\documentclass[aps,prx,twocolumn,superscriptaddress,amsmath,amssymb,10pt,epsfig,nofootinbib]{revtex4-2}
\usepackage{appendix}
\usepackage{amssymb, amsmath,amsthm}
\usepackage{graphicx}
\usepackage{verbatim}
\usepackage{color}
\usepackage{subcaption}
\usepackage{bbm}
\usepackage{caption}
\usepackage{tikz}
\usetikzlibrary{tikzmark}
\usepackage{pgflibraryarrows}
\usepackage{pgflibrarysnakes}
\usepackage{lipsum}
\usepackage{mathtools}
\usepackage{dirtytalk}
\usepackage{ae}
\usepackage[T1]{fontenc}
\usepackage[pdfstartview=FitV,linktocpage,colorlinks=true,linkcolor=darkblue,citecolor=darkblue,urlcolor=darkblue]{hyperref}
\usepackage{epsfig}
\usepackage{enumerate}
\usepackage{varioref}
\usepackage{cleveref}
\usepackage{makeidx}
\makeindex
\usepackage[english]{babel}
\usepackage[normalem]{ulem}
\usepackage{physics}
\usepackage{bbold}
\usepackage{float}
\usepackage{footnote}
\usepackage{gensymb}
\usepackage[version=4]{mhchem}
\usepackage[linesnumbered,ruled,vlined]{algorithm2e}

\newcommand{\beq}{\begin{equation}}
\newcommand{\eeq}{\end{equation}}
\newcommand{\beqq}{\begin{equation*}}
\newcommand{\eeqq}{\end{equation*}}
\newcommand\bea{\begin{array}}
\newcommand\eea{\end{array}}
\newcommand\beaa{\begin{array*}}
\newcommand\eeaa{\end{array*}}
\newcommand\beal{\begin{align}}
\newcommand\eeal{\end{align}}
\newcommand\beall{\begin{align*}}
\newcommand\eeall{\end{align*}}

\newcommand{\bphi}{\boldsymbol{\phi}}

\def\a{{\alpha}}

\def\s{\sigma}

\def\[{\left[}
\def\]{\right]}
\def\({\left(}
\def\){\right)}
\def\<{\langle}
\def\>{\rangle}

\definecolor{darkblue}{cmyk}{0.9,0.9,0,0}
\definecolor{greennote}{RGB}{0,135,41}

\begin{document}

\title{Strange metals and planckian transport in a gapless phase from spatially random interactions}

\author{Aavishkar A. Patel}
\email{apatel@flatironinstitute.org}
\affiliation{Center for Computational Quantum Physics, Flatiron Institute, 162 5th Avenue, New York, NY 10010, USA}
\author{Peter Lunts}
\email{plunts@fas.harvard.edu}
\affiliation{Department of Physics, Harvard University, Cambridge, MA 02138, USA}
\affiliation{Joint Quantum Institute and Department of Physics, University of Maryland, College Park, MD 20742, USA}
\author{Michael~S.~Albergo}
\email{malbergo@fas.harvard.edu}
\affiliation{Center for Cosmology and Particle Physics, New York University, New York, NY 10003, USA}
\affiliation{Courant Institute of Mathematical Sciences, New York University, New York, NY 10012, USA}
\affiliation{Society of Fellows, Harvard University, Cambridge, MA 02138, USA}

\date{\today}

\begin{abstract}

`Strange' metals that do not follow the predictions of Fermi liquid theory are prevalent in materials that feature superconductivity arising from electron interactions. In recent years, it has been hypothesized that spatial randomness in electron interactions must play a crucial role in strange metals for their hallmark linear-in-temperature ($T$) resistivity to survive down to low temperatures where phonon and Umklapp processes are ineffective, as is observed in experiments. However, a clear picture of how this happens has not yet been provided in a realistic model free from artificial constructions such as large-$N$ limits and replica tricks. We study a realistic model of two-dimensional metals with spatially random antiferromagnetic interactions in a non-perturbative regime, using numerically exact high-performance large-scale hybrid Monte Carlo and exact averages over the quenched spatial randomness. Our simulations reproduce strange metals' key experimental signature of linear-in-$T$ resistivity with a universal `planckian' transport scattering rate $\Gamma_\mathrm{tr} \sim k_B T/\hbar$ that is independent of coupling constants. We further find that strange metallicity in these systems is not associated with a quantum critical point, and instead arises from a phase of matter with gapless antiferromagnetic fluctuations that lacks long-range correlations and spans an extended region of parameter space: a feature that is also observed in several experiments. These gapless antiferromagnetic fluctuations take the form of spatially localized overdamped modes, whose presence could possibly be detected using recently developed nanoscale magnetometry methods. Our work paves the way for an eventual microscopic understanding of the role of spatial disorder in determining important properties of correlated electron materials. 

\end{abstract}

\maketitle

\section{Introduction}
\label{sec:intro}

Strange metals are compressible states of fermionic matter at finite density that defy expectations of the conventional Landau Fermi liquid (FL) theory of metals for the behavior of various observables, such as electrical resistivity and specific heat \cite{Chowdhury2022review}. They are present in many quasi-two dimensional (2D) correlated-electron superconductors at temperatures above the superconducting transition temperature, most notably the cuprates \cite{Legros2019}, as well as in heavy fermion compounds \cite{Nguyen2021} and other strongly correlated metallic systems \cite{Hayes2016, Jiang2023, NickelateSM, Jaoui2022, StrangeKagome}. A microscopic understanding of strange metals has long been believed to be a baseline for a microscopic understanding of correlated-electron superconductors, as the latter emerge from the former upon cooling.

The most striking feature of strange metals that violates the expectations from FL theory is their electrical resistivity ($\rho$), which has a linear dependence on temperature ($T$) that extends down to values of $T$ far below those that can give rise to such a dependence via conventional electron-phonon mechanisms in Fermi liquids. Additional features that are at odds with FL theory include larger specific heat capacities than those expected for Fermi liquids \cite{Girod2021}, and optical conductivities with linear-in-frequency ($\omega$) transport scattering rates \cite{Michon2022}. 

Quantum phase transitions (QPTs) are more or less always observed in materials that show strange metal behavior. Strange metals are often observed in a finite-temperature window (which is often termed a  `critical fan') above a putative quantum critical point (QCP) between two metallic phases of matter, one which possesses some kind of order and one which does not \cite{SachdevKeimer2011}. Furthermore, strange metals sometimes extend over an entire range of values of parameters controlling quantum phase transitions, instead of being confined to a critical fan \cite{Cooper2009, GreeneReview, Christos2022, NickelateSM, Jaoui2022}. These observations have led to many theoretical studies of models of metallic QCPs with various symmetry-breaking patterns or other kinds of order parameters \cite{Hertz1976, Millis1993}, in an attempt to eventually explain strange metals as the non-Fermi liquids that emerge when Fermi surfaces interact strongly with the fluctuating bosonic order parameters near criticality. 

A basic requirement for any viable model of strange metal behavior is the ability to produce a linear-in-$T$ electrical resistivity from electron-electron interactions alone, as phonons cannot be relied upon to do so at low values of $T$ below the Debye and Bloch–Grüneisen temperature scales. Without phonons, the only non-momentum conserving processes that can give rise to a nonzero electrical resistance in translationally-invariant lattice models of metals are Umklapp processes \cite{ZimanBook, holographicQM}. These Umklapp processes involve high energy excitations living at reciprocal lattice wavevectors that are not easily activated at low $T$, which gives rise to weak $T^2$ temperature dependencies of the electrical resistivity \cite{ZimanBook} that are incompatible with linear-in-$T$ strange metal behavior. Therefore, it was deduced \cite{holographicQM, Chowdhury2022review, Aldape2022, Patel2023universal} that strange metals must leverage microscopic randomness (disorder) in order to obtain sufficient levels of momentum relaxation for linear-in-$T$ resistivity at low $T$, and that translationally-invariant models of metallic QCPs are insufficient to explain strange metal behavior.       

Unlike the traditional description of disorder in textbooks on solid state physics, in which impurity potentials simply scatter electrons elastically \cite{ZimanBook}, disorder in strange metals must scatter electrons {\it inelastically} in order to be in agreement with experiments such as those of Refs. \cite{Michon2022, Chen2023}: Ref. \cite{Michon2022} found that the electron scattering rate in strange metals is strongly frequency dependent, and Ref. \cite{Chen2023} found that shot noise is strongly suppressed. These findings are incompatible with elastic electron-impurity scattering as well as with high-temperature electron-phonon scattering (which is also elastic) \cite{Li2024, Nikolaenko2023}, as they both produce frequency independent scattering rates and also no suppression of shot noise. Spatially non-uniform scattering of fermions off critical low-energy bosons, achieved by introducing quenched disorder into the Yukawa coupling between them, produces scattering that is {\it both} inelastic and non-momentum conserving, which is precisely the kind of scattering needed to achieve strange metal transport \cite{Aldape2022, Patel2023universal, Patel2023localization}. 

In this work, we therefore non-perturbatively investigate a realistic lattice model with such a disordered Yukawa coupling. We take a model of itinerant 2D fermions coupled to a fluctuating near-critical bosonic order parameter that represents commensurate collinear antiferromagnetic order. We numerically solve our lattice model exactly using a hybrid Monte Carlo (HMC) method \cite{DuaneHMC}, which was recently adapted for the study of metallic (and particularly quantum critical) systems \cite{Lunts2023HMC}, and applied to a sign-problem-free action that describes the non-disordered version of the aforementioned system \cite{Berg2012sign, Lunts2023HMC}. 

Relative to the non-disordered case, large spatial sizes are even more crucial to the disordered case that we consider here, as not only do we need to access long-wavelength fluctuations, but we also need to ensure that the system `self-averages' over disorder, {\it i.e.} that it does not end up being controlled by only a small portion of the disorder landscape. In addition to large spatial sizes, the method also needs to efficiently handle low temperatures and near-critical regimes in the phase diagram. We therefore enhance the capabilities of the HMC method via several technical improvements that allow for near-optimal performance whilst leveraging the massively parallel capabilities of modern graphics processing unit (GPU) hardware, thereby enabling the accurate extraction of low energy properties despite the enormous computational complexity of the problem. 

\begin{figure*}
    \includegraphics[width=0.98\textwidth]{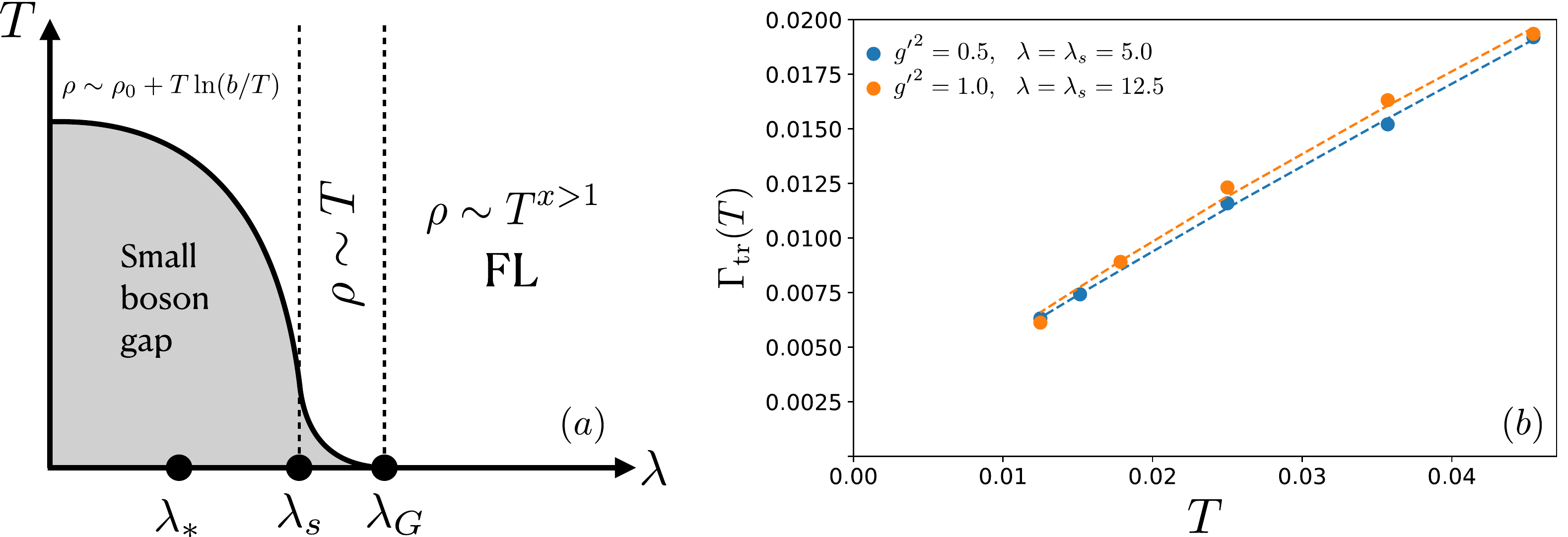}
    \caption{(a) A sketch of the phase diagram of our model with disordered Yukawa coupling, within the numerically accessible temperature window. A ``quantum Griffiths phase" with short-range correlations, but which  nevertheless hosts a gapless bosonic sector, onsets for $\lambda \leq \lambda_G$. A strange metal with MFL behavior and $T$-linear resistivity is obtained within this phase for $\lambda_s \leq \lambda \leq \lambda_G$, with the largest $T$-linear resistivity occuring at $\lambda = \lambda_s$. The strange metal is not associated with a QCP due to the lack of long-range correlations. For $\lambda < \lambda_s$, short-range order (SRO) manifests, which leads to a finite residual resistivity $\rho_0$. A logarithmic correction to the $T$-linear dependence of the resistivity also appears in this regime. As $\lambda$ is reduced below $\lambda_\ast$, a gradual crossover to long-range order starts to occur, with no sharp phase transition or QCP. For $\lambda > \lambda_G$, the bosonic sector is gapped and the fermions form a FL phase, which leads to a weaker $T$-dependence of the resistivity. (b) The transport scattering rate $\Gamma_\mathrm{tr}(T) \approx \alpha_0 T$ ($k_B = \hbar = 1$) at $\lambda = \lambda_s$ for two different values of the disordered Yukawa coupling ($g'$). The slope $\alpha_0 \approx 0.4$ is insensitive to $g'$.}
    \label{fig:phase_diag}
\end{figure*}

We find that the disorder strongly modifies bosonic properties at low energies, giving rise to overdamped and spatially localized bosonic modes. The fermions, in contrast, remain spatially extended. Relative to the non-disordered case, the disorder delays the onset of the long-range order in the bosonic sector that eventually must occur as the value of the transition tuning parameter ($\lambda$) is reduced, giving rise to a ``quantum Griffiths phase" {\cite{Patel2023localization, Hoyos2007, Vojta2009}} in which the boson modes are gapless but lack long-range correlations, and which spans a range of values of $\lambda$ between those that give rise to long-range order and those that give rise to a gapped bosonic phase. The fermions display marginal Fermi liquid (MFL) behavior \cite{VarmaMFL} and strange metallic $T$-linear resistivity over an extended range of values of $\lambda$ that comprises part of this ``quantum Griffiths phase", rather than at a QCP. These features are summarized in a phase diagram shown in Fig. \ref{fig:phase_diag} (a). 

Most strikingly, we find that the transport scattering rate ($\Gamma_\mathrm{tr}$) associated with the value of $\lambda$ that produces the largest slope of $T$-linear resistivity ($\lambda = \lambda_s$) is universal, {\it i.e.} $\Gamma_\mathrm{tr} (T)|_{\lambda=\lambda_s} \approx \alpha_0 k_B T/\hbar$, where $\alpha_0 \sim O(1)$ and is {\it independent} of the strength of the disordered Yukawa coupling (Fig. \ref{fig:phase_diag} (b)). This so-called ``planckian transport" is reminiscent of transport measurements on the cuprates, where it is also found that $\alpha_0 \sim O(1)$ and is nearly the same across several different compounds with varying degrees of disorder and electron interactions \cite{Legros2019, Bruin2013}, at the optimal doping that produces the largest slope of $T$-linear resistivity.

The rest of this paper is organized as follows: in Sec. \ref{sec:model}, we describe the model that we will exactly simulate. In Sec. \ref{sec:numerics} we analyze results for various physical quantities, with a specific focus on transport properties. In Sec. \ref{sec:hmc}, we summarize our technical improvements to the HMC algorithm. Finally, we discuss connections to experimental work and future directions in the field of correlated electron systems with disordered interactions in Sec. \ref{sec:discussion}.

\section{Spin-fermion model with disordered interactions}
\label{sec:model}

In this work, we consider the antiferromagnetic (AFM) ordering transition in 2D metals. AFM ordering is a ubiquitous property of the undoped cuprates, and AFM fluctuations are observed in many parts of the cuprate phase diagram, even relatively far from the AFM ordering transition \cite{Zhu2023}. Therefore, AFM criticality is generally believed to be of at least some relevance to the cuprates and their strange metal behavior, as well as to other classes of materials that exhibit both strange metal behavior and AFM ordering, such as the iron pnictides \cite{Hayes2016, Nakajima2020}. We will consider the effects of disordered interactions on the minimal theory describing the AFM ordering transition for our numerical simulations, which describes the onset of antiferromagnetic order at a finite wavevector $\mathbf{Q} = \mathbf{Q}_{\text{AF}}$, and is known in the literature as the `spin-fermion' model \cite{Abanov2003, Abanov2000, Metlitski2010, SchliefLuntsLee2d, LuntsSchliefLeeEps, SchliefLuntsLeeEps2, Berg2012sign, CarstenO32020, Lunts2023HMC}.  

\subsection{Action}

The `spin-fermion' model for the AFM ordering transition consists of fermions interacting with the bosonic order parameter for the transition, which is the collective staggered spin excitation. The sign-problem-free version of this model was given in Ref. \cite{Berg2012sign}, and we first write it down for the translationally-invariant (non-disordered) case: 

\begin{widetext}
\begin{equation}
\begin{split}
\mathcal{S}_0[\boldsymbol \phi, \psi, \psi^\dagger] & =  
\int d \tau
\sum_{\boldsymbol r, \boldsymbol r'} 
\sum_{\alpha=a,b}
\sum_{\sigma=\uparrow,\downarrow} 
\sum_{j=1}^{N_f}
\psi^\dagger_{\a,\s,j,\tau,\boldsymbol r}
\left[ (\partial_{\tau} - \mu_\alpha) \delta_{\boldsymbol r, \boldsymbol r'} 
- t_{\a, \boldsymbol r, \boldsymbol r'} \right] 
\psi_{\a,\s,j,\tau,\boldsymbol r'}
\\ & 
+
\int d \tau \sum_{\boldsymbol r}
\Bigl[
\frac{1}{2c^2} (\partial_\tau \boldsymbol{\phi}_{\tau,\boldsymbol r})^2 + \frac{1}{2} (\nabla \boldsymbol{\phi}_{\tau,\boldsymbol r})^2 + \frac{\lambda}{2} (\boldsymbol{\phi}_{\tau,\boldsymbol r})^2
+ \frac{u}{4} (\boldsymbol{\phi}_{\tau,\boldsymbol r} \cdot \boldsymbol{\phi}_{\tau,\boldsymbol r})^2
\Bigr]
\\ & 
+ g
\sum_{\s,\s'=\uparrow,\downarrow}
\sum_{j=1}^{N_f}
\int d \tau 
\sum_{\boldsymbol r}
e^{i \boldsymbol{Q}_{\text{AF}} \cdot \boldsymbol r} \;
\boldsymbol{\phi}_{\tau,\boldsymbol r} \cdot 
\Bigl[
\psi^\dagger_{a,\s,j,\tau,\boldsymbol r} \;
\boldsymbol{\tau}_{\s,\s'}  \;
\psi_{b, \s',j,\tau,\boldsymbol r}
+ \text{h.c.} \Bigr].
\end{split}
\label{eq:sign-problem-free action}
\end{equation}
\end{widetext}

In Eq. (\ref{eq:sign-problem-free action}), $\tau$ is imaginary time, $\boldsymbol{r} = (x, y)$ is the 2D spatial lattice coordinate of the square lattice. The $\psi_{\a,\s,j}$ are the fermions (electrons) which have a band index $\a \in \{ a,b \}$, a spin index $\s \in \{ \uparrow, \downarrow \}$, and an auxiliary flavor index $j = 1, \dots, N_f$ (the reason for the two bands and this additional degree of freedom are discussed below). $t_{\a, \boldsymbol r, \boldsymbol r'}$ are the real-space hopping amplitudes of the two bands, and the chemical potentials $\mu_\alpha = (-1)^\alpha\mu$ of the two bands are of opposite sign but equal magnitude. The field $\boldsymbol \phi$ is the bosonic three-component vector order parameter, and $\lambda$ is its bare `squared mass', which tunes the transition. The (translationally-invariant) Yukawa coupling between the boson and the fermion inter-band `spin' is denoted by $g$, and $\boldsymbol{\tau} = (\tau_x, \tau_y, \tau_z)$ is the vector of Pauli matrices. The wavevector $\mathbf{Q}_{\text{AF}} = (\pi,\pi)$ leads to commensurate AFM ordering. Also present is a self-interaction of the bosonic field, with coupling $u$.

The reason for the two-band construction is to avoid the sign problem in quantum Monte Carlo simulation, which is removed due to an anti-unitary fermion symmetry in the enlarged band-spin space \cite{Berg2012sign}. However, the two-band construction continues to reproduce the critical properties of the usual single-band minimal model. The additional $j$ index for the fermions and the corresponding \textit{even} number of flavors $N_f$ is necessary specifically for our HMC method (c.f. Sec. \ref{sec:hmc}). Its presence changes the numerical values of certain quantities in a predictable way, but does not qualitatively affect the physics of the system.  

\begin{figure*}
    \centering
    \includegraphics[width=0.98\textwidth]{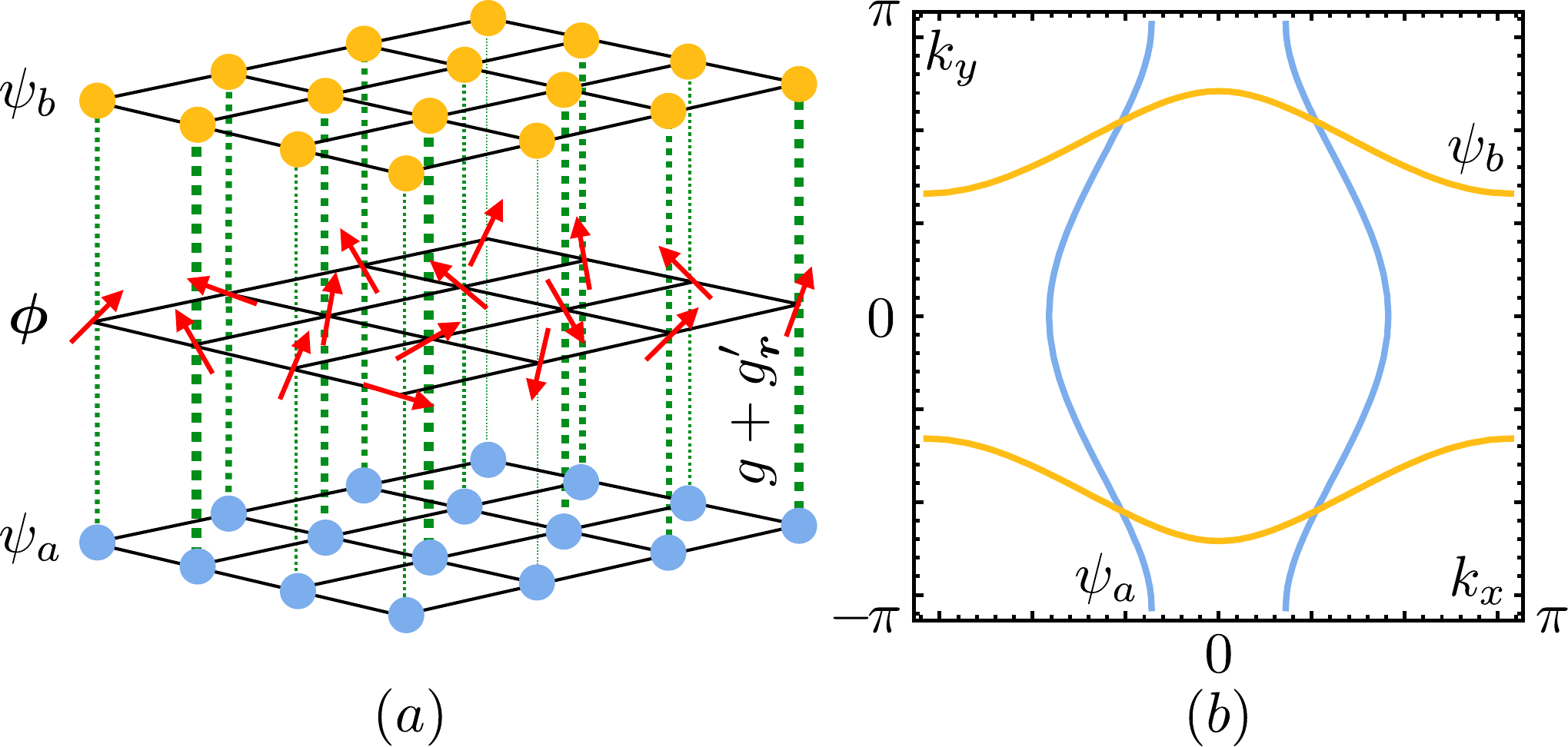}
    \caption{(a) A cartoon of the model that we simulate. It consists of two bands ($a,~b$) of fermions ($\psi$), mutually coupled via the bosonic antiferromagnetic order parameter field $\boldsymbol{\phi}$. The coupling $g + g'_{\boldsymbol{r}}$ fluctuates randomly in space. (b) The Fermi surface of the model (which is strictly defined only in the absence of disorder), showing the $a$ and $b$ bands. We use $t_{a,\boldsymbol{r}, \boldsymbol{r}'} = \delta_{\boldsymbol{r}',\boldsymbol{r}\pm\hat{x}} + 0.5 \times \delta_{\boldsymbol{r}',\boldsymbol{r}\pm\hat{y}}$, $t_{b,\boldsymbol{r}, \boldsymbol{r}'} = 0.5 \times \delta_{\boldsymbol{r}',\boldsymbol{r}\pm\hat{x}} + \delta_{\boldsymbol{r}',\boldsymbol{r}\pm\hat{y}}$, $\mu_a=-0.5$, and $\mu_b = 0.5$ throughout this work.}
    \label{fig:model}
\end{figure*}

Now we add spatial disorder, which can be introduced most generally in three channels: the interaction, chemical potential, and boson mass,
\begin{widetext}
\begin{equation}
\begin{split}
\mathcal{S}_{\mathrm{dis}}[\boldsymbol \phi, \psi, \psi^\dagger ] & =  
\int d \tau
\sum_{\boldsymbol r, \boldsymbol r'} 
\sum_{\alpha=a,b}
\sum_{\sigma=\uparrow,\downarrow} 
\sum_{j=1}^{N_f} 
\psi^\dagger_{\a,\s,j,\tau,\boldsymbol r}
\left[(\partial_{\tau} - \mu_{\alpha} - (-1)^{\alpha} \mu'_{\boldsymbol{r}}) \delta_{\boldsymbol r, \boldsymbol r'} 
- t_{\a, \boldsymbol r, \boldsymbol r'} \right] 
\psi_{\a,\s,j,\tau,\boldsymbol r'}
\\ & 
+
\int d \tau \sum_{\boldsymbol r}
\Bigl[
\frac{1}{2c^2} (\partial_\tau \boldsymbol{\phi}_{\tau,\boldsymbol r})^2 + \frac{1}{2} (\nabla \boldsymbol{\phi}_{\tau,\boldsymbol r})^2 + \frac{\lambda + \lambda'_{\boldsymbol r}}{2} (\boldsymbol{\phi}_{\tau,\boldsymbol r})^2
+ \frac{u}{4} (\boldsymbol{\phi}_{\tau,\boldsymbol r} \cdot \boldsymbol{\phi}_{\tau,\boldsymbol r})^2
\Bigr]
\\ & 
+
\sum_{\s,\s'=\uparrow,\downarrow}
\sum_{j=1}^{N_f}
\int d \tau 
\sum_{\boldsymbol r}
(g + g'_{\boldsymbol r})
e^{i \boldsymbol{Q}_{\text{AF}} \cdot \boldsymbol r} \;
\boldsymbol{\phi}_{\tau,\boldsymbol r} \cdot 
\Bigl[
\psi^\dagger_{a,\s,j,\tau,\boldsymbol r} \;
\boldsymbol{\tau}_{\s,\s'}  \;
\psi_{b, \s',j,\tau,\boldsymbol r}
+ \text{h.c.} \Bigr].
\end{split}
\label{eq:all disordered variables action}
\end{equation}
\end{widetext}
Here, $\mu'_{\boldsymbol{r}}, \lambda'_{\boldsymbol{r}}, g'_{\boldsymbol{r}}$ are spatially uncorrelated random variables drawn independently from the zero-mean truncated normal distributions 
\begin{equation}
\label{eq:disorder:dist}
\begin{split}
\mu'_{\mathbf{r}} \sim &\mathsf{TN}(0, {\mu'}^2, -2\mu', 2\mu'), \\
\lambda'_{\mathbf{r}} \sim &\mathsf{TN}(0, {\lambda'}^2, -2\lambda', 2\lambda'), \\
g'_{\mathbf{r}} \sim &\mathsf{TN}(0, {g'}^2, -2g', 2g'),    
\end{split}
\end{equation}
respectively, where $\mu',\lambda',g'$ determine the width of the distribution and the corresponding strength of disorder in each variable. A caricature of the model is shown in Fig. \ref{fig:model}.

In Eq. (\ref{eq:all disordered variables action}), we have shown the spatially random components in three different channels for generality. The on-site chemical potential $\mu'_{\boldsymbol{r}}$ is the most commonly studied disorder parameter in textbook solid state physics; however its role in transport is limited to elastic scattering that produces a mostly $T$-independent contribution to the residual resistivity \cite{ZimanBook}, and it does not directly contribute to strange metal behavior \cite{Patel2023universal}. We therefore do not consider it in our simulations. The disordered Yukawa interaction $g'_{\boldsymbol{r}}$ is what was argued in Ref. \cite{Patel2023universal} to be crucial for strange metal transport, by allowing the single-particle self energy to also manifest in the transport scattering rate. The boson mass disorder $\lambda'_{\boldsymbol{r}}$ is relevant at low energies \cite{Patel2023localization}, however in an exact simulation, disorder in the Yukawa interaction is expected to automatically generate disorder in the boson mass parameter via the boson self-energy; it is therefore sufficient to consider disorder solely in the Yukawa interaction in order to capture both its own effects as well as the effects of boson mass disorder. We therefore follow this recipe in setting up the model for our simulations.

Furthermore, in order to observe the largest effect of the randomness $g'_{\boldsymbol{r}}$ in the Yukawa interaction, we also set the average Yukawa interaction $g = 0$. This makes the total Yukawa interaction equally strong for scattering between any two momentum points on the Fermi surface (FS). In this limit, the physics therefore depends only on the fermion density of states and not the precise geometry of the FS, which allows us to test the effects of disordered interactions in the clearest way possible. Finally we set $N_f = 2$: as close to the single-flavor ($N_f = 1$) theory as possible while still keeping $N_f$ even to allow for simulation with our HMC method. The model we simulate is thus

\begin{widetext}
\begin{equation}
\begin{split}
\mathcal{S}[\boldsymbol \phi, \psi, \psi^\dagger] & = 
\int d \tau
\sum_{\boldsymbol r, \boldsymbol r'} 
\sum_{\alpha=a,b}
\sum_{\sigma=\uparrow,\downarrow} 
\sum_{j=1}^{2}
\psi^{\dagger}_{\a,\s,j,\tau,\boldsymbol r}
\left[ (\partial_{\tau} - \mu_{\alpha}) \delta_{\boldsymbol r, \boldsymbol r'} 
- t_{\a, \boldsymbol r, \boldsymbol r'} \right] 
\psi_{\a,\s,j,\tau,\boldsymbol r'}
\\ & 
+
\int d \tau \sum_{\boldsymbol r}
\Bigl[
\frac{1}{2c^2} (\partial_\tau \boldsymbol{\phi}_{\tau,\boldsymbol r})^2 + \frac{1}{2} (\nabla \boldsymbol{\phi}_{\tau,\boldsymbol r})^2 + \frac{\lambda}{2} (\boldsymbol{\phi}_{\tau,\boldsymbol r})^2
+ \frac{u}{4} (\boldsymbol{\phi}_{\tau,\boldsymbol r} \cdot \boldsymbol{\phi}_{\tau,\boldsymbol r})^2
\Bigr]
\\ & 
+
\sum_{\s,\s'=\uparrow,\downarrow}
\sum_{j=1}^{2}
\int d \tau 
\sum_{\boldsymbol r}
g'_{\boldsymbol r} \;
e^{i \boldsymbol{Q}_{\text{AF}} \cdot \boldsymbol r} \;
\boldsymbol{\phi}_{\tau,\boldsymbol r} \cdot 
\Bigl[
\psi^\dagger_{a,\s,j,\tau,\boldsymbol r} \;
\boldsymbol{\tau}_{\s,\s'}  \;
\psi_{b, \s',j,\tau,\boldsymbol r}
+ \text{h.c.} \Bigr].
\end{split}
\label{eq:action:g'}
\end{equation}

Measurements of an operator $\mathcal{O}$ of the full theory are then computed with a quenched disorder average $\overline{\langle\mathcal{O}\rangle}$:
\begin{equation}
\begin{split}
\overline{\langle \mathcal{O}\rangle} &= \frac{1}{N_d}\sum_{l=1}^{N_d} \langle\mathcal{O}\rangle_l,~~
\langle\mathcal{O}\rangle_l = \frac{\int \mathcal{D}\boldsymbol\phi \mathcal{D} \psi \mathcal{D} \psi^\dagger \; \mathcal{O} \, 
e^{-\mathcal S_l}}{\int \mathcal{D}\boldsymbol\phi \mathcal{D} \psi \mathcal{D} \psi^\dagger \;
e^{-\mathcal S_l}},
\end{split}
\label{eq:disorder:avg}
\end{equation}
where the index $l$ denotes a specific disorder realization drawn from the distributions in Eq. (\ref{eq:disorder:dist}), and $N_d$ is the number of disorder realizations used to compute the quenched disorder average. The corresponding statistical variance of the observables has contributions from both disorder and the Monte Carlo \cite{ballesteros1998a, Vojta2004}.
\end{widetext}

\section{Numerical results}
\label{sec:numerics}

We simulate the action of Eq. (\ref{eq:action:g'}) via the aforementioned HMC method. We choose a large value of $c=100$ to ensure that the boson's dynamics are solely due to its random Yukawa interactions with the fermions. We also set $u=0$ so that any effective boson self-interactions are also generated indirectly from the random Yukawa coupling. The parameters defining the fermion dispersion are given in the caption of Fig. \ref{fig:model}. We will consider two different values of the spread of the random Yukawa coupling, $g' = 1/\sqrt{2}$ and $g' = 1$, which correspond to interaction energy scales ${g'}^2 = E_F/5$ and ${g'}^2 = 2 E_F/5$ respectively, where $E_F = 5/2$ is the Fermi energy. Our results are for $g' = 1/\sqrt{2}$, unless otherwise noted. The fermion bandwidth is $12 E_F/5 = 6$. 

For each system we study, we combine results from $N_d = 10$ disorder realizations (unless mentioned otherwise), which is sufficient for the convergence of the measurements of our greatest interest. We simulate 2D systems of dimensions $L\times L$, with periodic boundary conditions. Unless otherwise noted, we use $L = 40$.

\subsection{Bosonic properties}
\label{sec:bosonic_properties}
\noindent 

\begin{figure}
\centering
    \includegraphics[width=0.48\textwidth]{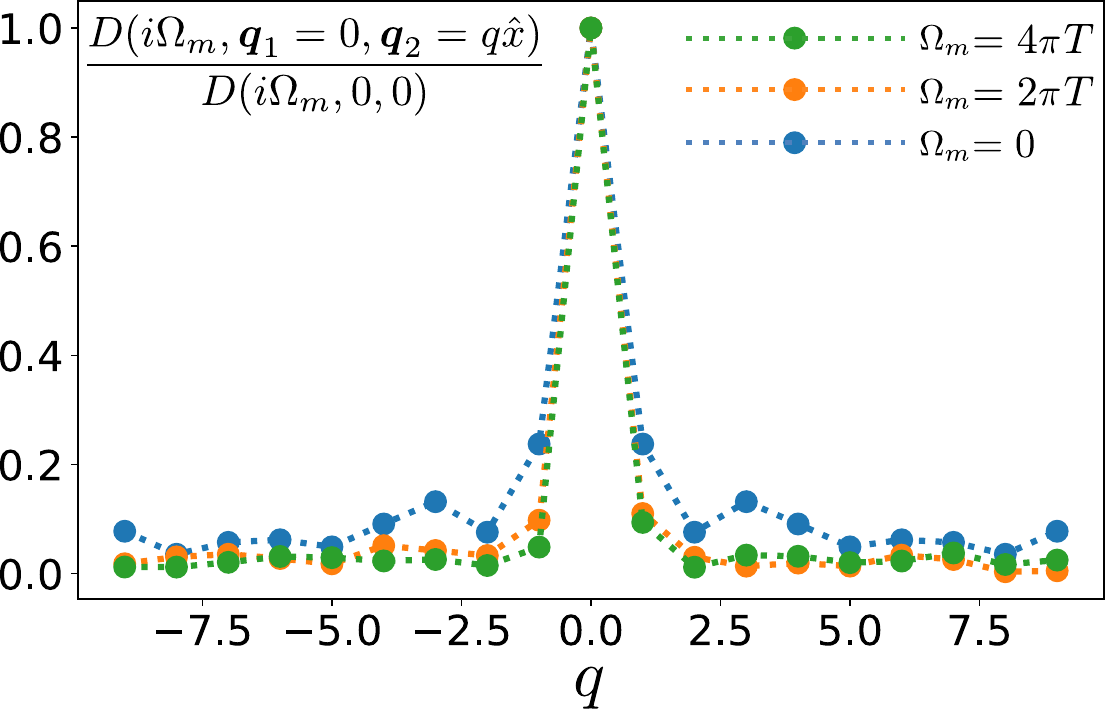}
    \caption{The off-diagonal-in-momentum boson propagator, measured for a single disorder realization with $\lambda = \lambda_s = 5.0 < \lambda_G$, at inverse temperature $\beta=66$. The zeroth Matsubara frequency shows a significant off-diagonal component.}
    \label{fig:OD_boson_propagator}
\end{figure}

In this section, we describe the static and dynamic properties of the boson $\boldsymbol{\phi}$ across the phase diagram shown in Fig. \ref{fig:phase_diag} (a). The features of our model are chiefly determined by the value of tuning parameter $\lambda$. The key to these features in the phase diagram, which differ significantly from those in the phase diagrams of similar non-disordered models \cite{BergQMCReview}, is the strong breaking of translational invariance at low energies in the bosonic sector. This is most pronounced in the ``quantum Griffiths phase" ($\lambda \leq \lambda_G$) in which the bosonic sector is gapless, and is easily visualized by examining the boson propagator in momentum space for a single disorder realization:
\begin{equation}
D(i\Omega_m, \boldsymbol{q}_1, \boldsymbol{q}_2) = \frac{1}{3}\langle \boldsymbol{\phi}^*_{i\Omega_m, \boldsymbol{q}_1} \cdot \boldsymbol{\phi}_{i\Omega_m, \boldsymbol{q}_2}\rangle,
\end{equation}
an example of which is shown in Fig. \ref{fig:OD_boson_propagator}. The momentum-off-diagonal components of $D(i\Omega_m, \boldsymbol{q}_1, \boldsymbol{q}_2 \neq \boldsymbol{q}_1)$ become significant at small bosonic Matsubara frequencies $\Omega_m = 2m\pi T$, and are largest at $\Omega_m=0$, signaling strong disorder in the bosonic sector at low energies.

\begin{figure*}
\centering
    \includegraphics[width=0.98\textwidth] {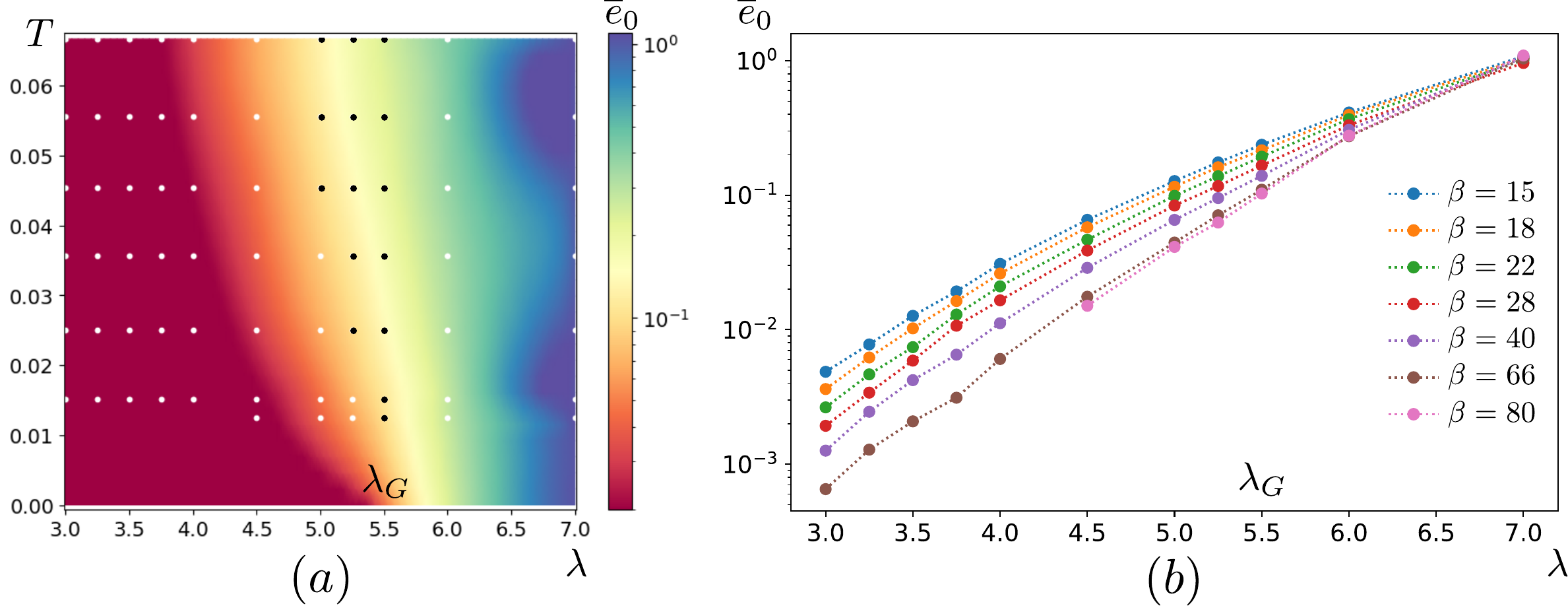}   
    \caption{(a) The disorder-averaged boson gap $\bar{e}_0$ as a function of $\lambda$ and $T$. The dots (both black and white) correspond to the actual data points, with the values in between them generated by cubic interpolation. The values for $T < 1/80$ ($T < 1/66$ for $\lambda < 4.5$)  are obtained using the extrapolation procedure in Appendix \ref{app:boson_gap_susc}. The colors saturate to the respective endpoints of the color bar when $\overline{e}_0$ goes out of range. The red area is part of the ``small boson gap region" of Fig. \ref{fig:phase_diag}(a). (b) $\overline{e}_0$ as a function of $\lambda$ for different inverse temperatures $\beta$, showing an exponential decrease as $\lambda$ is reduced in the ``quantum Griffiths phase" ($\lambda < \lambda_G$).}
    \label{fig:mass_gap}
\end{figure*}

Due to the strong disorder in the bosonic sector, determining the boson gap (and thereby the extent of the gapless ``quantum Griffiths phase") requires diagonalizing the static boson propagator for a given disorder realization:
\begin{equation}
\begin{split}
    D(i\Omega_m=0, \boldsymbol{r}_1, \boldsymbol{r}_2) &=
    \frac{1}{3}
    \langle\boldsymbol{\phi}_{i\Omega_m = 0, \boldsymbol{r}_1}\cdot\boldsymbol{\phi}_{i\Omega_m = 0,\boldsymbol{r}_2}\rangle \\
    &= \sum_{\a = 0}^{L^2 - 1} \frac{\eta_{\a,\boldsymbol{r}_1} \eta_{\a,\boldsymbol{r}_2}}{e_{\a}},
\end{split}
\end{equation}
in terms of eigenvalues of the inverse static boson propagator, $e_{\a}$, and eigenvectors, $\eta_{\a,\boldsymbol{r}}$. The boson gap is then given by the lowest eigenvalue $e_0$. We show the disorder-averaged boson gap for different $T$ and $\lambda$ in Fig. \ref{fig:mass_gap}. This is computed as the inverse of the average of the largest eigenvalue of the static boson propagator over disorder realizations:
\begin{equation}
\overline{e}_0 = \left[\frac{1}{N_d}\sum_{l=1}^{N_d}\left(\frac{1}{e_0}\right)_l\right]^{-1}. 
\end{equation}
In the ``quantum Griffiths phase" ($\lambda \leq \lambda_G$), $\overline{e}_0$ decreases exponentially as $\lambda$ is reduced. In Appendix \ref{app:boson_gap_susc}, we provide additional analysis of the $T$ dependence of $\overline{e}_0$, showing that $\overline{e}_0(T\rightarrow 0) = 0$ for $\lambda < \lambda_G$, as is expected for a gapless phase.    

\begin{figure*}
\centering
    \includegraphics[width=0.98\textwidth]{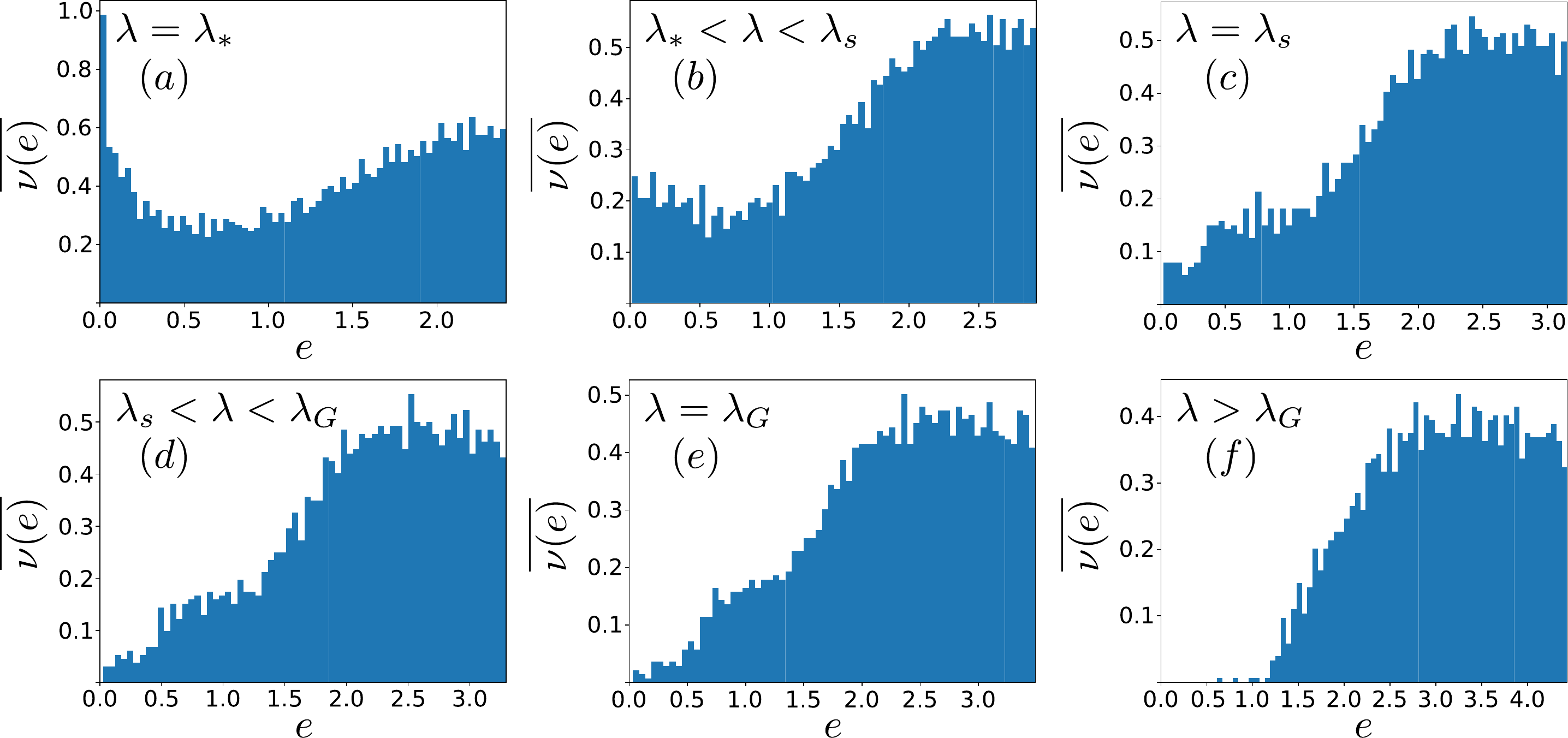}
    \caption{Disorder-averaged boson density of states $\overline{\nu(e)}$ for inverse temperature $\beta = 66$. The $\lambda$ values are (a) - (f): $\lambda = 3.5,~4.5,~5.0,~5.25,~5.5,~7.0$ respectively. For $\lambda \leq \lambda_G = 5.5$ the boson is gapless, with $\overline{\nu(e \to 0)} \neq 0$. For $\lambda < \lambda_s = 5.0$, $\overline{\nu(e)}$ shows an upturn at small $e$.}
    \label{fig:boson_DOS}
\end{figure*}

We further analyze the spectral properties of the bosonic sector through the disorder-averaged density of states $\overline{\nu(e)}$ of the eigenvalues $e$, $\overline{\nu(e)} = (1/N_d)\sum_{l=1}^{N_d} \nu_l(e)$, shown in Fig. \ref{fig:boson_DOS}. In the gapless ``quantum Griffiths phase", $\overline{\nu(e \rightarrow 0)} \neq 0$ as expected. On the other hand,  $\overline{\nu(e \rightarrow 0)} = 0$ for $\lambda > \lambda_G$. For $\lambda < \lambda_s$, $\overline{\nu(e)}$ shows an upturn at small $e$.

\begin{figure*}
\centering
    \includegraphics[width=0.98\textwidth]{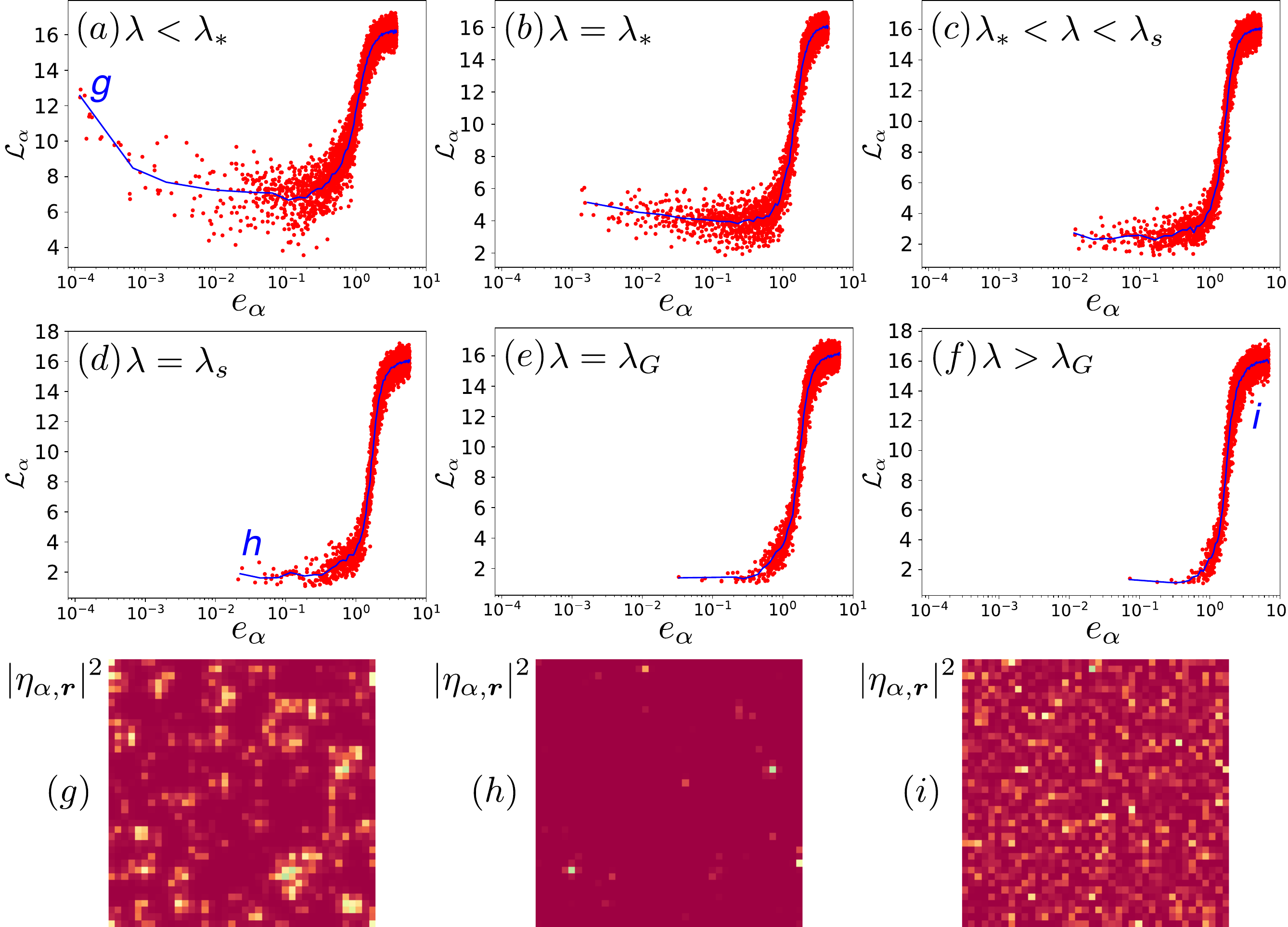}
    \caption{Localization lengths $\mathcal{L}_\alpha$ of the eigenvectors $\eta_{\alpha,\boldsymbol{r}}$ of the inverse static boson propagator plotted versus their corresponding eigenvalues $e_\alpha$, at inverse temperature $\beta = 66$. The blue curves are the average values of $\mathcal{L}_\alpha$ within $e_\alpha$ bins. The $\lambda$ values are (a) - (f): $\lambda = 2.5,~3.5,~4.5,~5.0,~5.5,~6.0$ respectively. (g) - (i): The density profiles $|\eta_{\alpha,\boldsymbol{r}}|^2$ of eigenvectors of the three different kinds discussed in the main text.}
    \label{fig:LLs_and_eigenmodes}
\end{figure*}

We now turn to the structure of the eigenvectors $\eta_{\alpha, \mathbf{r}}$ that describe the spatial profiles of the bosonic modes. Due to the strong disorder, these modes are not plane waves, and many of them are spatially localized. We therefore characterize them using localization lengths $\mathcal{L}_{\alpha}$, which are defined using the inverse participation ratio $\mathcal{I}_{\alpha}$:
\begin{equation}
    \mathcal{I}_{\alpha} = \sum_{\boldsymbol{r}} \abs{\eta_{\alpha, \boldsymbol{r}}}^4, \quad \mathcal{L}_{\alpha} = \frac{1}{\sqrt{2 \mathcal{I}_{\alpha}}}.
\end{equation}
The localization lengths $\mathcal{L}_\alpha$ are shown in Fig. \ref{fig:LLs_and_eigenmodes} for various values of $\lambda$, along with the density profiles $|\eta_{\alpha, \boldsymbol{r}}|^2$ of selected eigenvectors. 

We see that there are three distinct types of eigenvectors \cite{Patel2023localization}: (i) at large eigenvalues $e_\alpha \gg 1$, we have superpositions of plane waves with large $\mathcal{L}_{\alpha} \sim O(L/2)$ (Fig. \ref{fig:LLs_and_eigenmodes} (i)). (ii) For $\lambda > \lambda_\ast$ and $e_\alpha < 1$, the eigenvectors are tightly localized, with $\mathcal{L}_\alpha$ of the order of a few lattice spacings (Fig. \ref{fig:LLs_and_eigenmodes} (h)). As we subsequently show, these low-energy localized bosonic modes are overdamped. They are therefore similar in spirit to the localized ``two-level systems" \cite{Bashan23, Bashan24} that have been conjectured to be a source of electron scattering in strange metals. However, they are not postulated degrees of freedom like in Refs. \cite{Bashan23, Bashan24}, and instead emerge naturally from a realistic model at low energies. Their density increases substantially as $\lambda$ is reduced below $\lambda_G$ and the gapless ``quantum Griffiths phase" is entered. (iii) Finally, for $\lambda \leq \lambda_\ast$, the localized modes begin to coalesce into extended states at the lowest values of $e_\alpha$ (Fig. \ref{fig:LLs_and_eigenmodes} (g)), leading to a slow increase in $\mathcal{L}_\alpha$ as $e_\alpha$ is decreased, with $\mathcal{L}_\alpha \sim [\ln(1/e_\alpha)]^p$. This is qualitatively similar to the ``activated scaling" noted in quantum Griffiths phases arising in random transverse-field Ising models \cite{FisherIsing1992, FisherIsing1995, MotrunichIsing2000}. 

The localization length of the low-lying eigenvectors is expected to increase to $\mathcal{L}_{\alpha} \sim O(L/2)$ for $\lambda \ll \lambda_\ast$, leading to the asymptotic establishment of long-range order (LRO) as $\lambda \rightarrow -\infty$. An analysis of the disorder-averaged uniform static bosonic susceptibility $\chi$ however reveals the lack of LRO within the numerically accessible window of $\lambda$; on the other hand, analysis of the corresponding dynamic susceptibility $\chi(i\Omega_m)$ shows that the bosonic order parameter develops a frozen-in-time component for $\lambda < \lambda_s$, leading to short-range order (SRO) in that regime (Appendix \ref{app:boson_gap_susc}). As discussed further in Appendix \ref{app:boson_gap_susc}, this SRO is glassy in nature, leading to long autocorrelation times and metastability in the Monte Carlo dynamics. The glassiness likely arises from the random-in-sign long range couplings in the effective boson action obtained after integrating out the fermions, and is reminiscent of ``cluster glass" phases \cite{VladGlass1, VladGlass2} proposed earlier in the literature, that are believed to preempt phase transitions and QCPs between quantum Griffiths phases and LRO phases in itinerant electron systems.

\begin{figure*}
\centering
    \includegraphics[width=0.98\textwidth]{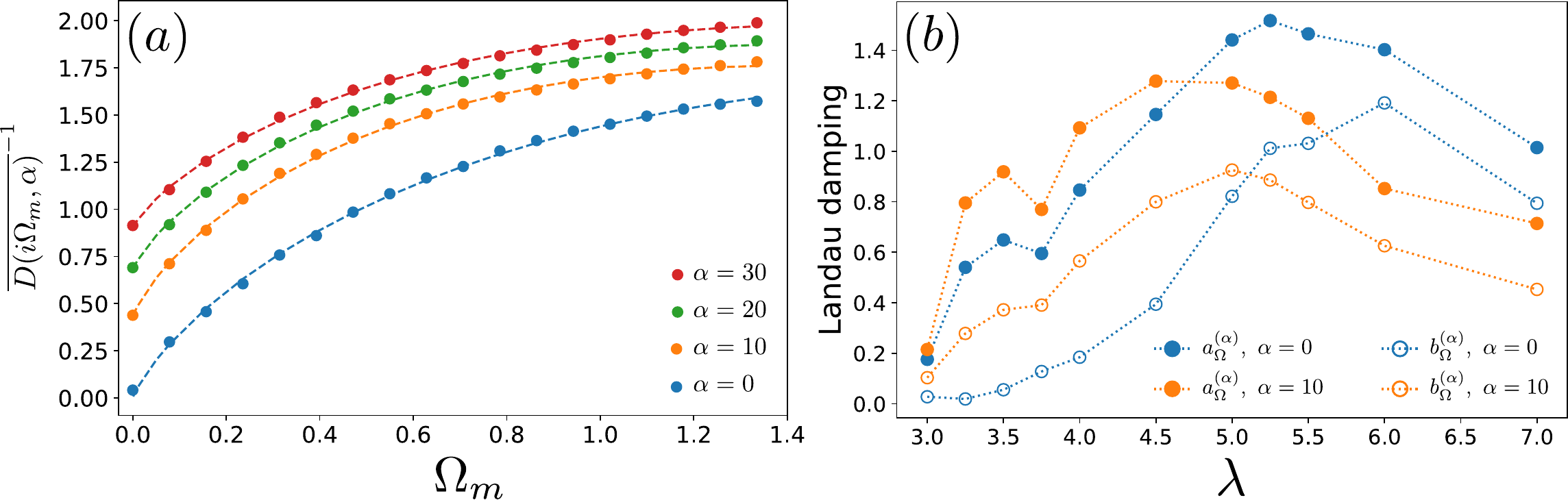}
    \caption{(a) The inverse disorder-averaged boson propagator in the eigenbasis, $\overline{D(i\Omega_m, \a)}^{-1}$, measured for $\lambda = \lambda_s =  5.0$ and inverse temperature $\beta = 80$, plotted as a function as a function of $\Omega_m$. Other values of $\lambda$ behave similarly and are shown in Appendix \ref{app:more_plots}. The $\Omega_m$ dependence fits well to $a_\Omega^{(\alpha)}\abs{\Omega_m} + b_\Omega^{(\alpha)} \abs{\Omega_m} \log(1/\abs{\Omega_m}) + \overline{e}_{\alpha}$ (dashed curves). (b) The dependence of the coefficients $a_\Omega^{(\alpha)},~b_\Omega^{(\alpha)}$ on $\lambda$ at $\beta = 66$.}
    \label{fig:eigenbasis_boson_propagator}
\end{figure*}

To investigate the dynamics of the localized modes, we compute the Matsubara frequency dependence of the disorder-average of the diagonalized boson propagator, $\overline{D(\alpha)}=(1/N_d)\sum_{l=1}^{N_d}D_l(i\Omega_m, \alpha)$, where $\alpha$ is the eigenvalue index that is held fixed across all the disorder realizations involved. In Fig. \ref{fig:eigenbasis_boson_propagator} (a) we show $\overline{D(i\Omega_m,\alpha)}^{-1}$ as a function of $\Omega_m$ for values of $\alpha$ that correspond to the low-energy localized modes (Fig. \ref{fig:LLs_and_eigenmodes} (h)). We find that $\overline{D(i\Omega_m,\alpha)}^{-1}$ is a non-analytic function of $\Omega_m$ that is very different from the $\sim \Omega_m^2$ dependence expected for free bosons, indicating strong Landau damping that arises from coupling of the bosons to the fermions. The $\Omega_m$ dependence is very well fit by a logarithmically corrected Ohmic Landau damping term, $\overline{D(i\Omega_m,\alpha)}^{-1} \approx f_{(\alpha)}(i\Omega_m) + \overline{e}_{\alpha}$, where $\overline{e}_\alpha = [(1/N_d)\sum_{l=1}^{N_d}(1/e_\alpha)_l]^{-1}$, and $f_{(\alpha)}(i\Omega_m) = a_{\Omega}^{(\alpha)}\abs{\Omega_m} + b_{\Omega}^{(\alpha)} \abs{\Omega_m} \log(1/\abs{\Omega_m})$ at all values of $\lambda$ and $T$ that we consider.

\begin{figure*}
\centering
    \includegraphics[width=0.98\textwidth]{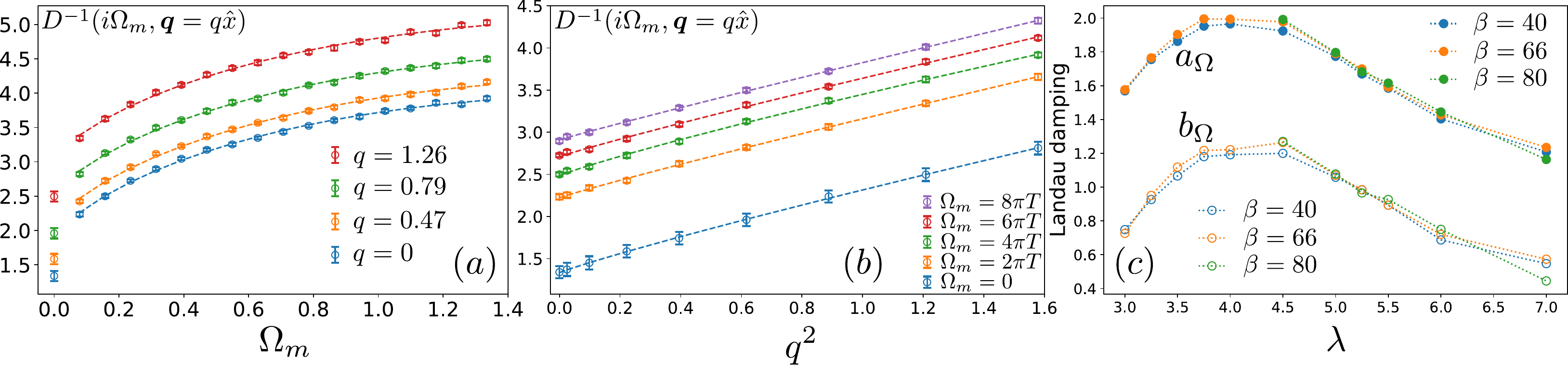}
    \caption{The inverse disorder-averaged boson propagator in momentum space, $D^{-1}(i\Omega_m, \boldsymbol{q})$, measured for $\lambda = \lambda_s =  5.0$ and inverse temperature $\beta = 80$, plotted as a function of (a) $\Omega_m$ and (b) $q^2$ in momentum space. Other values of $\lambda$ are shown in Appendix \ref{app:more_plots}, and behave similarly. All dependencies (excluding the $\Omega_m = 0$ points in panel (a)) fit well to $a_x\abs{x} + b_x \abs{x} \log(1/\abs{x}) + c_x$ (dashed curves), with $x = \Omega_m$ or $x = q^2$. For both the $\Omega_m$ and $q^2$ dependencies, the coefficients $a_x,~b_x$ are largely constant as a function of the non-varied variable (except at $\Omega_m = 0$), which leads to the additively separable form of $D^{-1}(i\Omega_m > 0, \boldsymbol{q})$ mentioned in the main text. (c) The dependence of the coefficients $a_\Omega,~b_\Omega$ (averaged over the various momentum values, since no momentum stands out), as a function of $\lambda$ for various values of $\beta$. The coefficients are rather insensitive to $\beta$.}
    \label{fig:diag_boson_propagator}
\end{figure*}

The disorder-averaged boson propagator in momentum space $\overline{D(i\Omega_m, \boldsymbol{q}, \boldsymbol{q}')} = \delta_{\boldsymbol{q},\boldsymbol{q}'}D(i\Omega_m, \boldsymbol{q})$ is a translationally invariant (momentum-diagonal) quantity, that contains information about scaling properties typically used to characterize translationally invariant systems. We show $D^{-1}(i\Omega_m,\boldsymbol{q})$ as a function of both $\Omega_m$ (Fig. \ref{fig:diag_boson_propagator} (a)) and $\boldsymbol{q}$ (Fig. \ref{fig:diag_boson_propagator} (b)). At all values of $\lambda$ and $T$ that we consider, we find that $D^{-1}(i\Omega_m > 0, \boldsymbol{q}) \approx f_\Omega(i\Omega_m) + f_q(\boldsymbol{q}) + c_D$, where $f_\Omega(i\Omega_m) = a_{\Omega}\abs{\Omega_m} + b_{\Omega} \abs{\Omega_m} \ln(1/\abs{\Omega_m})$, and $f_q(\boldsymbol{q}) = a_{q}\abs{\boldsymbol{q}}^2 + b_{q} \abs{\boldsymbol{q}}^2 \ln(1/\abs{\boldsymbol{q}}^2)$. This implies an average relative scaling of $\Omega_m \sim \abs{\boldsymbol{q}}^2$ at long wavelengths and intermediate-to-low energies ($2\pi T \lesssim \Omega \ll E_f$), which is observed, for instance, in neutron scattering measurements of the dynamical spin structure factor in the cuprates \cite{Hayden2023}. While this scaling does imply a dynamical critical exponent of $z=2$ at such energy scales, for $\lambda \leq \lambda_*$ the ``activated scaling" discussed earlier implies a value of $z=\infty$ at the lowest energies \cite{FisherIsing1992}. We also see that $D^{-1}(i\Omega_m = 0, \boldsymbol{q}) \approx f^0_q(\boldsymbol{q}) + \chi^{-1}$, with $f^0_q(\boldsymbol{q}) = a^0_q |\boldsymbol{q}|^2 + b_q^0|\boldsymbol{q}|^2\ln(1/\abs{\boldsymbol{q}}^2) < f_q(\boldsymbol{q})$, and $\chi^{-1} < c_D$. The discontinuity between the zeroth and finite Matsubara frequencies is another signature of the strong disorder in the boson sector at low energies.

\subsection{Fermionic properties}

We now turn to investigating how the bosonic properties discussed previously affect the fermions, focusing first on the single-particle properties of the fermions, and then on the charge transport properties of the system (which arise entirely from the fermion sector). Of particular interest to us are the single-particle and transport scattering rates. These give us key insight into the nature of the metallic phase, and can be directly compared to experimental observations. 

\subsubsection{Fermion occupation functions}

\begin{figure*}
    \centering
    \includegraphics[width=0.98\textwidth]{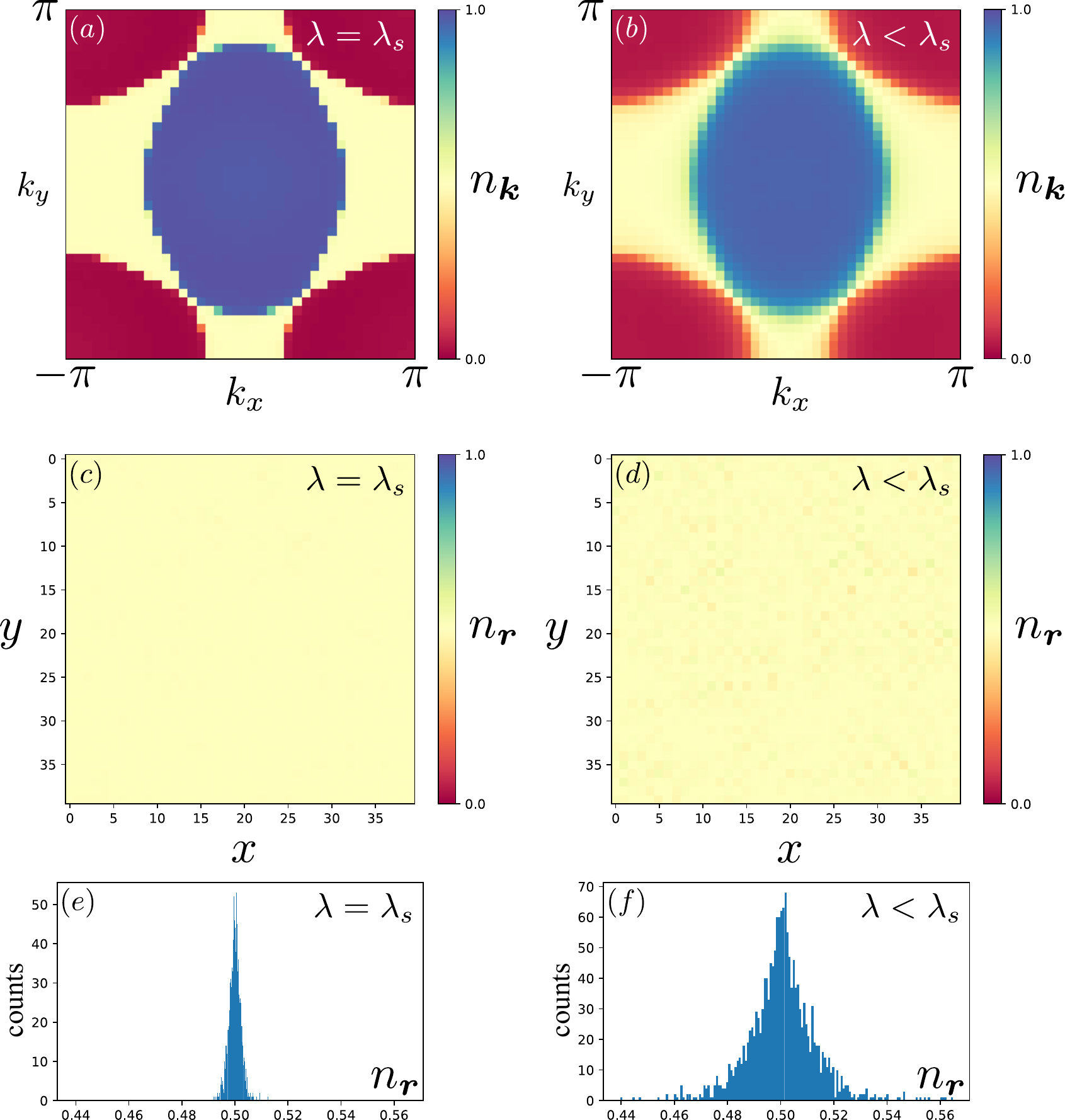}
    \caption{(a), (b) Fermion occupation functions $n_{\boldsymbol{k}}$ in momentum space. (c), (d) Fermion occupation functions $n_{\boldsymbol{r}}$ in real space. (e), (f) Histograms of $n_{\boldsymbol{r}}$. In the left column $\lambda = \lambda_s = 5.0$ (strange metal), and $\lambda = 2.5 < \lambda_s$ in the right column. All measurements are for a single disorder realization at inverse temperature $\beta = 66$. The strange metal (left column) has a sharp FS in momentum space (the momentum space occupancy in (a) follows from the FS structure shown in Fig. \ref{fig:model} (b)), and a highly homogeneous density of fermions in real space. For $\lambda < \lambda_s$ (right column), the FS becomes fuzzy and inhomogeneity starts to appear in the real space density.}
    \label{fig:fermion_occs}
\end{figure*}

We begin with studying the normalized fermion occupation functions in real ($\boldsymbol{r}$) and momentum ($\boldsymbol{k}$) space:
\begin{equation}
n_{\boldsymbol{r}}  = \frac{1}{8}\sum_{\alpha=a,b}\sum_{\sigma=\uparrow,\downarrow}\sum_{j=1}^2\langle \psi^\dagger_{\alpha,\sigma,j,\boldsymbol{r}} \psi_{\alpha,\sigma,j,\boldsymbol{r}} \rangle,
\end{equation}
\begin{equation}
n_{\boldsymbol{k}}  = \frac{1}{8}\sum_{\alpha=a,b}\sum_{\sigma=\uparrow,\downarrow}\sum_{j=1}^2\langle \psi^\dagger_{\alpha,\sigma,j,\boldsymbol{k}} \psi_{\alpha,\sigma,j,\boldsymbol{k}} \rangle,
\end{equation}
for a single disorder realization. These are shown in Fig. \ref{fig:fermion_occs}. We find that, despite strong disorder and localization in the bosonic sector, the strange metal ($\rho \sim T$) regime of Fig. \ref{fig:phase_diag} (a) has a sharp FS and no visible localization of fermions. On the other hand, for $\lambda < \lambda_s$ the FS becomes increasingly fuzzy as $\lambda$ is reduced: this is due to the onset of a residual (zero-energy) scattering rate of fermions in the momentum basis, as we shall demonstrate subsequently. As we argue later, this residual scattering rate is caused by a mechanism similar to an effective random potential. Therefore, the conventional 2D weak localization that is caused by random potentials \cite{LeeRamakrishnan} is expected to occur for $\lambda < \lambda_s$. However, due to the fact that the disordered interaction is attractive, we expect a pairing instability to set in at a much higher temperature than that at which localization could be observed. 

\subsubsection{Fermion self-energy}
\label{sec:sigma}

\begin{figure}
    \centering
    \includegraphics[width=0.48\textwidth]{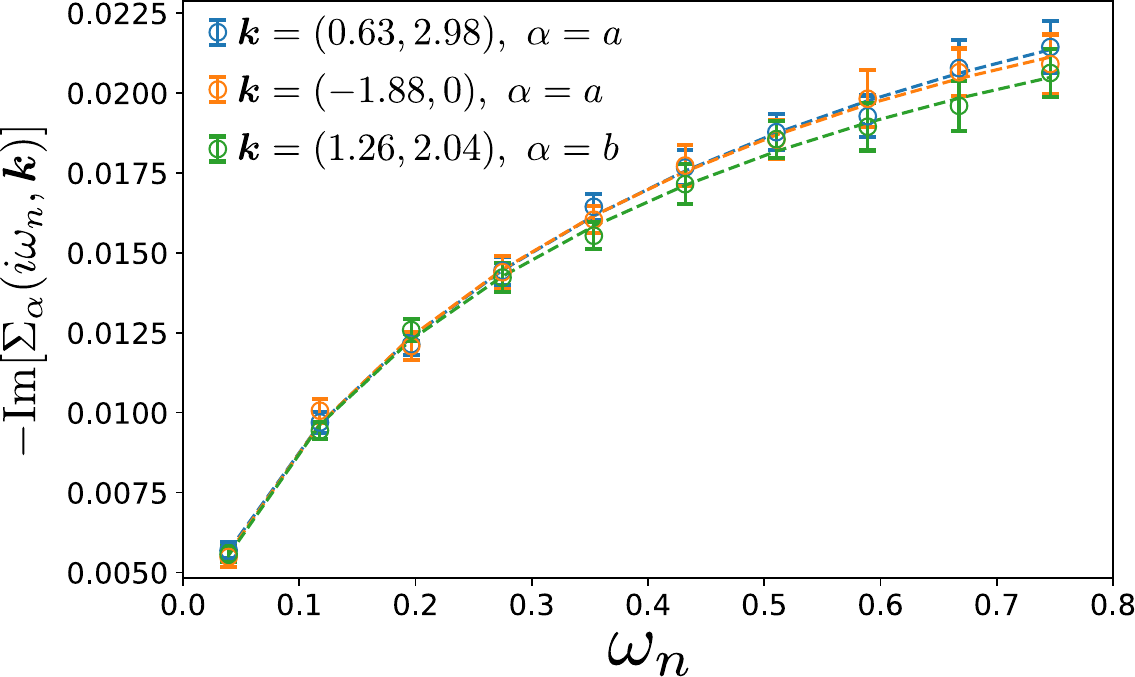} 
    \caption{The imaginary part of the Matsubara fermion self-energy, $-\Im[\Sigma_{\alpha}(i\omega_n, \boldsymbol{k})]$, measured for $\lambda = \lambda_s = 5.0$ and inverse temperature $\beta = 80$ at three different points on the FS, along with the corresponding MFL fits (dashed curves, see text). The curves are the same within error bars.}
    \label{fig:Sigma_multiple_k}
\end{figure}

The fermion self-energy, $\Sigma_{\alpha}(i\omega_n, \boldsymbol{k})$ (where $\omega_n = 2(n+1/2)\pi T$ are fermionic Matsubara frequencies) characterizes the dynamics of the low-energy fermion excitations that live near the FS. It can be defined using the disorder-averaged fermion Green's function $G_{\alpha}(i\omega_n, \boldsymbol{k})$, where 
\begin{equation}
\frac{1}{4}\sum_{\sigma=\uparrow,\downarrow}\sum_{j=1}^2\overline{\langle\psi_{\alpha,\sigma,j,i\omega_n,\boldsymbol{k}}\psi^\dagger_{\alpha,\sigma,j,i\omega_n, \boldsymbol{k}'}\rangle} = \delta_{\boldsymbol{k}, \boldsymbol{k}'} G_{\alpha}(i\omega_n, \boldsymbol{k}),
\end{equation} 
as
\begin{equation}
\Sigma(i\omega_n, \boldsymbol{k}) = G_{0\alpha}^{-1}(i\omega_n, \boldsymbol{k}) - G_{\alpha}^{-1}(i\omega_n, \boldsymbol{k})
\end{equation}
($G_{0\alpha}(i\omega_n, \boldsymbol{k})$ is the non-interacting fermion Green's function). Since the Yukawa interaction is entirely random in space, with no preferred direction or momentum scale, the self-energy is momentum-independent along the FS, as we show in Fig. \ref{fig:Sigma_multiple_k}. We therefore compute its average value over the FS,
\begin{equation}
\Sigma_{\text{FS}}(i\omega_n) \equiv \frac{1}{N_{\text{FS}}} \sum_{(\boldsymbol{k},\alpha) \in {\text{FS}}} \Sigma_{\alpha}(i\omega_n, \boldsymbol{k}),    
\end{equation}
which greatly reduces statistical noise in the self-energy measurement.

\begin{figure*}
    \centering
    \includegraphics[width=0.98\textwidth]{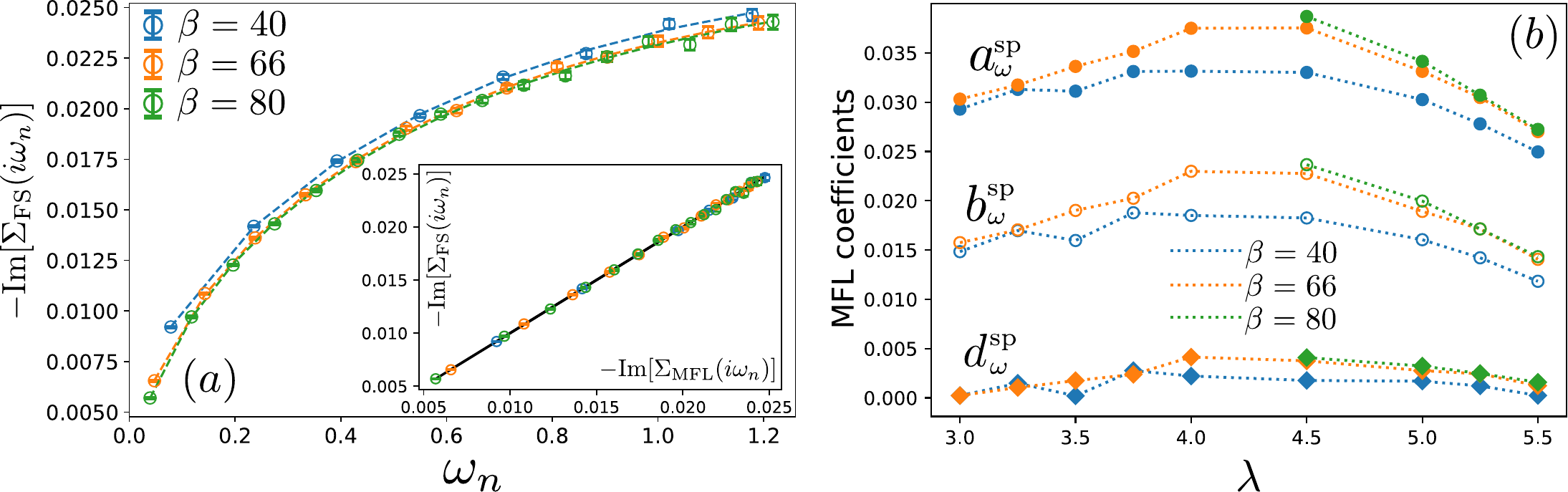}
    \caption{(a) The imaginary part of Matsubara fermion self-energy averaged over the FS, $-\text{Im}[\Sigma_\text{FS}(i\omega_n)]$, at $\lambda = \lambda_s = 5.0$ and different values of inverse temperature $\beta$, with fits to Eq. (\ref{eq:Sigma_MFL_fit}) (dashed curves and inset). The points in the inset lie on the $y = x$ line, indicating an excellent fit. Other values of $\lambda \leq \lambda_G$ behave similarly and are shown in Appendix \ref{app:more_plots}. (b) The dependence of the fit coefficients $a^{\text{sp}}_\omega, b^{\text{sp}}_\omega, d^{\text{sp}}_\omega$ of Eq. (\ref{eq:Sigma_MFL_fit}) on $\lambda$ and $\beta$.}
    \label{fig:self_energy}
\end{figure*}

We plot $\Sigma_{\text{FS}}(i\omega_n)$ in Fig. \ref{fig:self_energy} (a). As the real part $\text{Re}[\Sigma_{\text{FS}}(i\omega_n)] \approx 0$ (not shown), we only display the imaginary component. For $\lambda \leq \lambda_G$, we expect the coupling of fermions to gapless bosonic modes to destroy fermionic quasiparticles \cite{SachdevQPMbook}: consequently, in the frequency-driven scaling regime of $\Lambda_f \gg |\omega_n| \gg T$ ($\Lambda_f$ is an ultraviolet cutoff) $-\text{Im}[\Sigma_\text{FS}(\Lambda_f \gg |i\omega_n| \gg T)]$ should scale sub-linearly in $|\omega_n|$. We find that this is indeed the case, and $-\text{Im}[\Sigma_\text{FS}(\Lambda_f \gg |i\omega_n| \gg T)]$ fits perfectly to a MFL form:
\begin{align}
\label{eq:Sigma_MFL_fit}
-\text{Im}[\Sigma_{\text{MFL}}(i\omega_n)] & = a^{\text{sp}}_\omega~\omega_n + b^{\text{sp}}_\omega~\omega_n \ln(0.5/\abs{\omega_n})
\\ & + d^{\text{sp}}_\omega~\omega_n \ln^2(0.5/\abs{\omega_n}) + c^{\text{sp}}_\omega~\mathrm{sgn}(\omega_n) \nonumber
\end{align}
(Fig. \ref{fig:self_energy} (a)). This MFL form differs from the traditional one in Ref. \cite{VarmaMFL} by an additional logarithmic correction $d^{\text{sp}}_\omega~\omega_n \ln^2(0.5/\abs{\omega_n})$, which allows for a slightly stronger destruction of quasiparticles. We also add the coefficient $c^{\text{sp}}_\omega$ to allow for a finite $-\text{Im}[\Sigma_\text{FS}(i\omega_n = 0)]$ that arises due to the presence of thermal fluctuations at $T > 0$, as well as residual scattering rates that arise for $\lambda < \lambda_s$.

We now turn to investigating the behavior of the DC (zero frequency) single-particle scattering rate $\Gamma_{\text{sp}} = -\text{Im}[\Sigma_{\text{FS}}(i\omega_n = 0)]$. Since the lowest value of $|\omega_n|$ at finite $T$ is $\pi T$, obtaining this quantity requires extrapolating the self-energy at finite Matsubara frequencies, $-\text{Im}[\Sigma_{\text{FS}}(i\omega_n)]$, to $\omega_n = 0$. As is well known, correlation functions are analytic in $\omega_n$ in the temperature-driven scaling regime of $\Lambda_f \gg T \gtrsim |\omega_n|$ \cite{Sachdev_QPT}. Therefore, the non-analytic function in Eq. (\ref{eq:Sigma_MFL_fit}) cannot be used to perform the extrapolation to zero frequency. A common technique used instead is to perform the extrapolation using a polynomial spline fit to the finite-frequency data, that produces an analytic function as $\omega_n\rightarrow 0$ (see, {\it e.g.}, Refs. \cite{Fournier2024, WuTremblay2022, DumitrescuGeorgesPlanckian2022}). This procedure generally agrees well with more sophisticated analytic continuation methods for metallic systems \cite{Fournier2024}.

\begin{figure*}
\centering
    \includegraphics[width=0.98\textwidth]{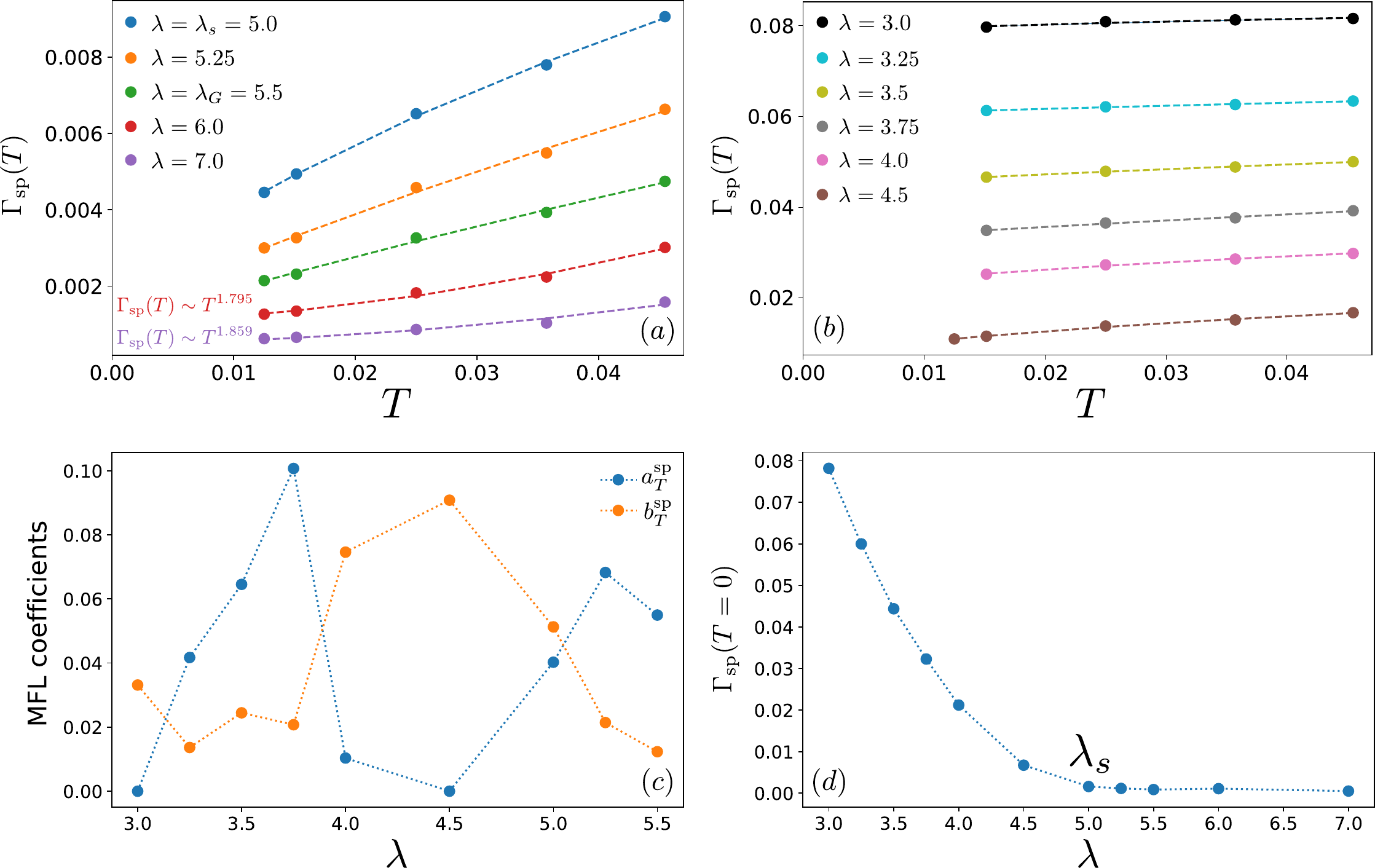}
    \caption{(a), (b) The DC single-particle scattering rate $\Gamma_{\text{sp}}$ for various values of $T$ and $\lambda$. The dashed curves are fits to Eqs. (\ref{eq:Gamma_sp_FL}, \ref{eq:Gamma_sp_MFL}) in their applicable regimes. (c) The dependence of the fit coefficients $a^{\text{sp}}_T,~b^{\text{sp}}_T$ of Eq. (\ref{eq:Gamma_sp_MFL}) on $\lambda$. (d) The residual single particle scattering rate $\Gamma_{\text{sp}}(T = 0)$ as a function of $\lambda$, inferred from Eqs. (\ref{eq:Gamma_sp_FL}, \ref{eq:Gamma_sp_MFL}) in their applicable regimes.}
    \label{fig:Gamma_sp}
\end{figure*}

We therefore use a cubic spline function with smoothing factor $s = 1$ to perform the extrapolation\footnote{Specificially, we use \href{https://docs.scipy.org/doc/scipy/reference/generated/scipy.interpolate.UnivariateSpline.html}{\texttt{scipy.interpolate.UnivariateSpline}}, and fit to Matsubara frequencies up to $|\omega_n| \approx E_F/2$.}. The resulting DC single-particle scattering rate is shown in Fig. \ref{fig:Gamma_sp}. For $\lambda > \lambda_G$, where bosonic excitations are gapped, $\Gamma_{\text{sp}}(T)$ is concave up, and fits to a power-law $T$-dependence 
\begin{equation}
\Gamma_{\text{sp}}^{\text{FL}}(T) = a^{\text{sp}}_T~T^{x > 1} + \Gamma_{\text{sp}}(T = 0),
\label{eq:Gamma_sp_FL}
\end{equation}
with $\Gamma_{\text{sp}}(T = 0) \approx 0$ (Fig. \ref{fig:Gamma_sp} (a)). This is unremarkable, as coupling to gapped bosons does not destroy fermion quasiparticles at low energies, and the DC scattering rates are therefore parametrically smaller than the thermal quasiparticle excitation energies (which are of order $T$).

For $\lambda \leq \lambda_G$, where the bosons are gapless, $\Gamma_{\text{sp}}(T)$ fits well to 
\begin{equation}
\Gamma_{\text{sp}}^{\text{MFL}}(T) = a^{\text{sp}}_T~T + b^{\text{sp}}_T~T\ln(0.5/T) + \Gamma_{\text{sp}}(T = 0)
\label{eq:Gamma_sp_MFL}
\end{equation}
(Fig. \ref{fig:Gamma_sp} (a), (b)). We deduce this fit Ansatz by analytically continuing Eq. (\ref{eq:Sigma_MFL_fit}) to real frequencies ($i\omega_n \rightarrow \omega + i0^+$): this gives a finite-frequency scattering rate $-\text{Im}[\Sigma^R_\text{MFL}(\omega)] = \pi b^{\text{sp}}_\omega |\omega|/2 + \pi d^{\text{sp}}_\omega |\omega| \ln(0.5/|\omega|) + c^{\text{sp}}_\omega$. Under the assumption of $\omega/T$ scaling, that is usually present in the dissipative dynamics of gapless systems \cite{Sachdev_QPT}, the $|\omega|$ dependence of the finite-frequency scattering rate then translates to the $T$ dependence of our Ansatz. 

The residual scattering rate $\Gamma_{\text{sp}}(T = 0)$ determined by our fits is essentially zero for $\lambda \geq \lambda_s$ but becomes nonzero for $\lambda < \lambda_s$, increasing with decreasing $\lambda$ (Fig. \ref{fig:Gamma_sp} (d)). The $\lambda$ dependence of $\Gamma_{\text{sp}}(T = 0)$ correlates strongly with the $\lambda$ dependence of the frozen component $\chi_0$ of the dynamic boson susceptibility $\chi(i\Omega_m)$ (see Appendix \ref{app:boson_gap_susc} and Fig. \ref{fig:boson_dynamic_susc} (d)), indicating that the residual scattering rate arises from the order parameter freezing and glassy SRO that occurs for $\lambda < \lambda_s$. This connection becomes clear upon examining the Yukawa interaction term in Eq. (\ref{eq:action:g'}): upon partial freezing of the order parameter fields $\boldsymbol{\phi}_{\tau,\boldsymbol{r}} \rightarrow \boldsymbol{\phi}^{\text{frz}}_{\boldsymbol{r}} + \boldsymbol{\phi}^{\text{dyn}}_{\tau, \boldsymbol{r}}$, the Yukawa interaction term transforms into
\begin{align}
\label{eq:freezing_mechanism}
& g'_{\boldsymbol r} \;
e^{i \boldsymbol{Q}_{\text{AF}} \cdot \boldsymbol r} \;
\boldsymbol{\phi}_{\tau,\boldsymbol r} \cdot 
\psi^\dagger_{a,\s,j,\tau,\boldsymbol r} \;
\boldsymbol{\tau}_{\s,\s'}  \;
\psi_{b, \s',j,\tau,\boldsymbol r} \\
& \rightarrow 
\boldsymbol{v}_{\boldsymbol{r}} \cdot 
\Bigl[
\psi^\dagger_{a,\s,j,\tau,\boldsymbol r} \;
\boldsymbol{\tau}_{\s,\s'}  \;
\psi_{b, \s',j,\tau,\boldsymbol r}
+ \text{h.c.} \Bigr] \nonumber \\
& + 
g'_{\boldsymbol r} \;
e^{i \boldsymbol{Q}_{\text{AF}} \cdot \boldsymbol r} \;
\boldsymbol{\phi}^{\text{dyn}}_{\tau,\boldsymbol r} \cdot 
\Bigl[
\psi^\dagger_{a,\s,j,\tau,\boldsymbol r} \;
\boldsymbol{\tau}_{\s,\s'}  \;
\psi_{b, \s',j,\tau,\boldsymbol r}
+ \text{h.c.} \Bigr]. \nonumber
\end{align}
The term proportional to $\boldsymbol{v}_{\boldsymbol{r}} = g'_{\boldsymbol{r}}e^{i\boldsymbol{Q}_{\text{AF}}\cdot \boldsymbol{r}}\boldsymbol{\phi}^{\text{frz}}_{\boldsymbol{r}}$ (that arises from the frozen component $\boldsymbol{\phi}^{\text{frz}}_{\boldsymbol{r}}$) is similar to a static random impurity potential for the fermions (albeit one in a spin-dependent and inter-band channel), which is the textbook mechanism for obtaining a residual scattering rate for both single-particle dynamics and collective transport \cite{ZimanBook, LeeRamakrishnan}. The remaining fluctuating component $\boldsymbol{\phi}^{\text{dyn}}_{\tau,\boldsymbol{r}}$ is gapless and allows for the MFL $T$-dependence of $\Gamma_{\text{sp}}$ to persist for $\lambda < \lambda_s$.

\subsubsection{Transport properties}
\label{sec:sigma_tr}

We turn to the main observable of interest, which is the electrical conductivity $\sigma$. We first compute the conductivity on the Matsubara frequency axis, $\sigma(i\Omega_m = 2\pi i m T)$ using the Kubo formula: 
\begin{equation}
\sigma(i\Omega_m) = \frac{1}{2|\Omega_m|}\left[\overline{\langle K\rangle} - \overline{\langle \boldsymbol{J}_{i\Omega_m}\cdot\boldsymbol{J}_{-i\Omega_m}\rangle}\right],
\end{equation}
where $\boldsymbol{J}_{i\Omega_m} = (J^x_{i\Omega_m}, J^y_{i\Omega_m})$ is the total current operator at Matsubara frequency $\Omega_m$ generated by the fermion hopping terms in Eq. (\ref{eq:action:g'}), and $K$ is the part of the fermion Hamiltonian comprised of the fermion hopping terms ({\it i.e.} the fermion kinetic energy) \cite{ScalapinoWhiteZhang1993}. 

We then parameterize the conductivity as
\begin{align}
\sigma(i\Omega_m) = \frac{W}{|\Omega_m| + \Sigma_{\text{tr}}(i\Omega_m)},
\label{eq:Drude}
\end{align}
which is simply the standard finite-frequency Drude formula for the complex conductivity (such as Eqs. (2), (3) of Ref. \cite{Michon2022}),
\begin{align}
\sigma(\Omega) & =  \frac{W}{-i\Omega + \Sigma^R_{\text{tr}}(\Omega)} \nonumber \\
& =  \frac{W}{-i\Omega\left(\frac{m^*_{\mathrm{tr}}(\Omega)}{m_{\mathrm{tr}}}\right)  + \frac{1}{\tau_{\mathrm{tr}}(\Omega)}},
\end{align}
analytically continued to the Matsubara frequency axis. Here, $\pi W$ is the free Drude weight that we compute exactly by applying the Kubo formula to the non-interacting and non-disordered system with $g'=0$. The quantity $\Sigma_{\text{tr}}(i\Omega_m)$ defines the ``transport self-energy" on the Matsubara frequency axis; for real frequencies, its real part $\text{Re}[\Sigma^R_{\text{tr}}(\Omega)]$ provides the familiar frequency-dependent transport scattering rate $1/\tau_{\mathrm{tr}}(\Omega)$. Consequently, a sensible result for $\Sigma_{\text{tr}}(i\Omega_m)$, measured indirectly as $W\sigma^{-1}(i\Omega_m) - |\Omega_m|$, implies the validity of a Drude picture of transport with a frequency-dependent scattering rate. We expect such a picture to be valid due to the isotropy of the fermion self-energy on the FS, which implies that all carriers are scattered equally strongly just like in the Drude model.

\begin{figure*}
    \centering
    \includegraphics[width=0.98\textwidth]{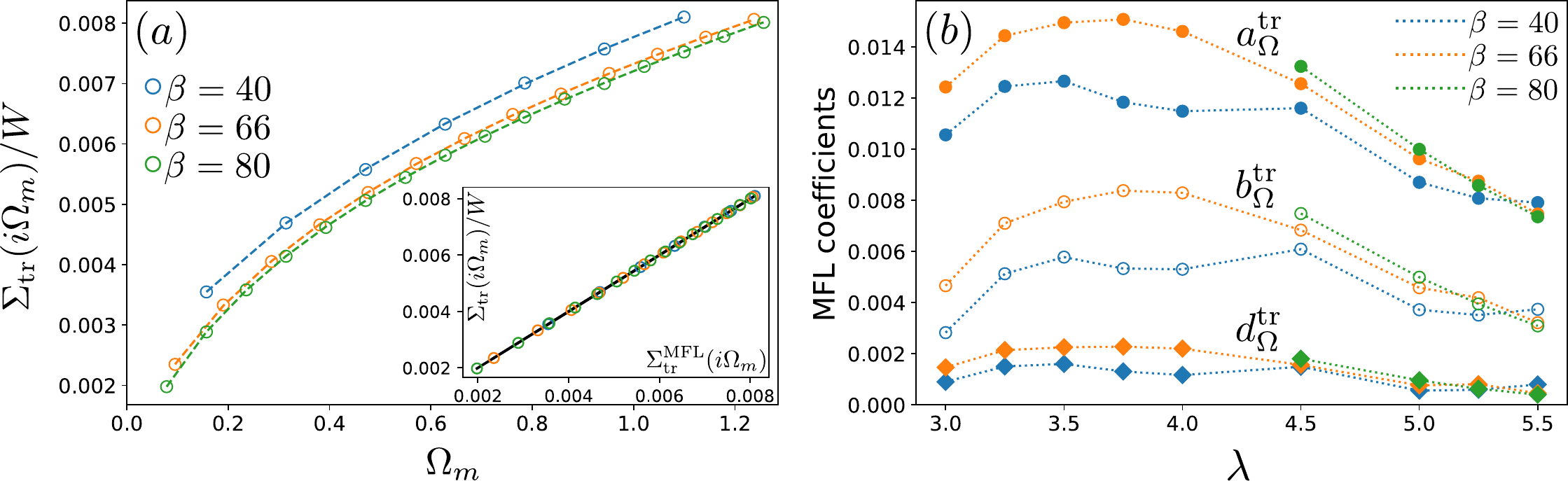}
    \caption{(a) The ``transport self-energy" $\Sigma_\text{tr}(i\Omega_m)$ on the Matsubara frequency axis, normalized by $(1/\pi)$ times  the free Drude weight, at $\lambda = \lambda_s = 5.0$ and different values of inverse temperature $\beta$ and with fits to Eq. (\ref{eq:Sigma_tr_MFL_fit}) (dashed curves and inset). The points in the inset lie on the $y = x$ line, indicating an excellent fit. Other values of $\lambda \leq \lambda_G$ behave similarly and are shown in Appendix \ref{app:more_plots}. (b) The dependence of the fit coefficients $a^{\text{tr}}_\Omega, b^{\text{tr}}_\Omega, d^{\text{tr}}_\Omega$ of Eq. (\ref{eq:Sigma_tr_MFL_fit}) on $\lambda$ and $\beta$.}
    \label{fig:transport_self_energy}
\end{figure*}

Upon extracting $\Sigma_{\text{tr}}(i\Omega_m)$ as described above, we indeed find sensible metallic behavior, with $\Sigma_{\text{tr}}(\Lambda_f \gg |i\Omega_m|) > 0$ and increasing as a function of $|\Omega_m|$. For $\lambda \leq \lambda_G$, where the bosonic modes are gapless, we find that $\Sigma_{\text{tr}}(\Lambda_f \gg |i\Omega_m|)/W$ fits perfectly to a MFL form that is just like Eq. (\ref{eq:Sigma_MFL_fit}) for the imaginary part of the fermion self-energy:
\begin{align}
\label{eq:Sigma_tr_MFL_fit}
\Sigma^{\text{MFL}}_{\text{tr}}(i\Omega_m) & = a^{\text{tr}}_\Omega~|\Omega_m| + b^{\text{tr}}_\Omega~|\Omega_m| \ln(0.5/\abs{\Omega_m})
\\ & + d^{\text{tr}}_\Omega~|\Omega_m| \ln^2(0.5/\abs{\Omega_m}) + c^{\text{tr}}_\Omega \nonumber
\end{align}
(Fig. \ref{fig:transport_self_energy} (a)). 

\begin{figure*}
\centering
    \includegraphics[width=0.98\textwidth]{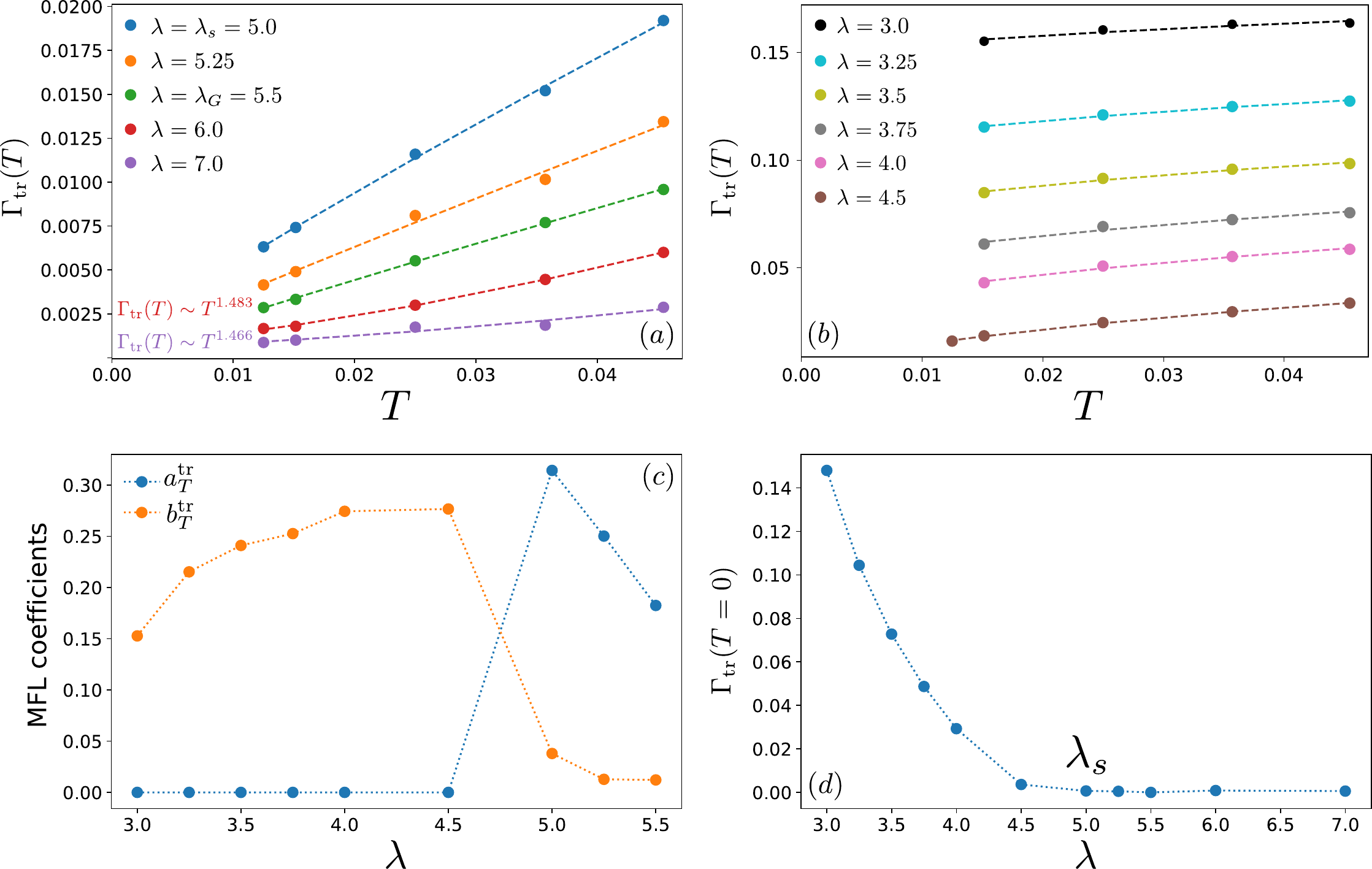}
    \caption{(a), (b) The DC transport scattering rate $\Gamma_{\text{tr}}$ for various values of $T$ and $\lambda$. The dashed curves are fits to Eqs. (\ref{eq:Gamma_tr_FL}, \ref{eq:Gamma_tr_MFL}) in their applicable regimes. (c) The dependence of the fit coefficients $a^{\text{tr}}_T,~b^{\text{tr}}_T$ of Eq. (\ref{eq:Gamma_tr_MFL}) on $\lambda$. (d) The residual transport scattering rate $\Gamma_{\text{tr}}(T = 0)$ as a function of $\lambda$, inferred from Eqs. (\ref{eq:Gamma_tr_FL}, \ref{eq:Gamma_tr_MFL}) in their applicable regimes.}
    \label{fig:Gamma_tr}
\end{figure*}

To make contact with the pi{\`e}ce de r{\'e}sistance of strange metals, we now compute the DC transport scattering rate $\Gamma_{\text{tr}} = \text{Re}[\Sigma_{\text{tr}}(i\Omega_m = 0)]$. The DC electrical resistivity is given by $\rho = \Gamma_{\text{tr}}/W$. We employ the same procedure that uses smoothed cubic splines to extrapolate to zero frequency as in Sec. \ref{sec:sigma}, the results of which are shown in Fig. \ref{fig:Gamma_tr}. For $\lambda > \lambda_G$, where the bosons are gapped, $\Gamma_{\text{tr}}(T)$ is concave up just like $\Gamma_{\text{sp}}(T)$, and fits to
\begin{equation}
\Gamma_{\text{tr}}^{\text{FL}}(T) = a^{\text{tr}}_T~T^{x > 1} + \Gamma_{\text{tr}}(T = 0),
\label{eq:Gamma_tr_FL}
\end{equation}
with $\Gamma_{\text{tr}}(T = 0) \approx 0$ (Fig. \ref{fig:Gamma_tr} (a)). This is again expected on similar grounds as for $\Gamma_{\text{sp}}(T)$, due to the existence of coherent fermion quasiparticles in that regime.

For $\lambda \leq \lambda_G$ the fermion quasiparticles get destroyed by coupling to gapless bosons, and $\Gamma_{\text{tr}}(T)$ is no longer concave up. Analytically continuing Eq. (\ref{eq:Sigma_tr_MFL_fit}) to real frequencies, and assuming that the $\omega/T$ scaling holds, we arrive at the same MFL Ansatz for $\Gamma_{\text{tr}}(T)$ as we have for $\Gamma_{\text{sp}}(T)$ (Eq. (\ref{eq:Gamma_tr_MFL})) in this regime:
\begin{equation}
\Gamma_{\text{tr}}^{\text{MFL}}(T) = a^{\text{tr}}_T~T + b^{\text{tr}}_T~T\ln(0.5/T) + \Gamma_{\text{tr}}(T = 0),
\label{eq:Gamma_tr_MFL}
\end{equation}
which fits the data well (Fig. \ref{fig:Gamma_tr} (a), (b)).

While $\Gamma_{\text{tr}}^{\text{MFL}}(T)$ is a mixture of purely linear and logarithmically corrected $T$-dependent contributions, they are seemingly mutually exclusive (Fig. \ref{fig:Gamma_tr} (c)). For $\lambda_s \leq \lambda \leq \lambda_G$, $\Gamma_{\text{tr}}(T)$ (and hence $\rho$) is almost exactly linear in $T$, giving rise to a strange metal (consequently $b_T^\text{tr} \approx 0$). On the other hand, the $\lambda < \lambda_s$ region is dominated by the logarithmic correction ($a_T^{\text{tr}} \approx 0$ and $b_T^\text{tr} > 0$), and does not have $\rho \sim T$ strange metal behavior. In the linear-in-$T$ regime, $\Gamma_{\text{tr}}(T)$ has a slope whose average value over the range of $T$ that we consider is $\alpha_0(\lambda = \lambda_s) \approx 0.4$ and $\alpha_0(\lambda = \lambda_G) \approx 0.2$. The maximal $T$-linear transport scattering rate occurs at $\lambda = \lambda_s$.

The residual transport scattering rate $\Gamma_{\text{tr}}(T = 0)$ is negligible for $\lambda \geq \lambda_s$, and then increases with a similar $\lambda$-profile as $\Gamma_{\text{sp}}(T = 0)$ (Fig. \ref{fig:Gamma_tr} (d)). This is expected, as the partial freezing mechanism of Eq. (\ref{eq:freezing_mechanism}) generates an effective static random potential, which has the same effect on transport scattering rates as it does on single-particle ones.  

\subsection{Universality of strange metal behavior}
\label{sec:universality}

In this subsection, we compare the DC scattering rates $\Gamma_{\text{tr}}$ and $\Gamma_{\text{sp}}$ already discussed at $g' = 1/\sqrt{2}$ with those at a larger value of the strength of the random Yukawa coupling ($g' = 1$). In addition to the independence on $g'$ of $\Gamma_{\text{tr}}(T)$ at $\lambda = \lambda_s$ discussed in Sec. \ref{sec:intro}, we also establish the same independence for $\Gamma_{\text{sp}}(T)$. Furthermore, we show that the scattering rates at different values of $g'$ can be unified in terms of a shifted and rescaled tuning parameter 
\begin{equation}
\tilde{\lambda} = \frac{\lambda - \lambda_s}{{g'}^2},    
\end{equation} 
thereby extending the planckian transport noted in Sec. \ref{sec:intro} for $\lambda = \lambda_s$ to the entirety of the strange metal phase with $\lambda_s \leq \lambda \leq \lambda_G$. Finally, by analyzing the scaling of the low-energy boson propagator with $g'$, we provide a mechanistic explanation for the universality of the scattering rates.

\begin{figure*}
\centering
    \includegraphics[width=0.98\textwidth]{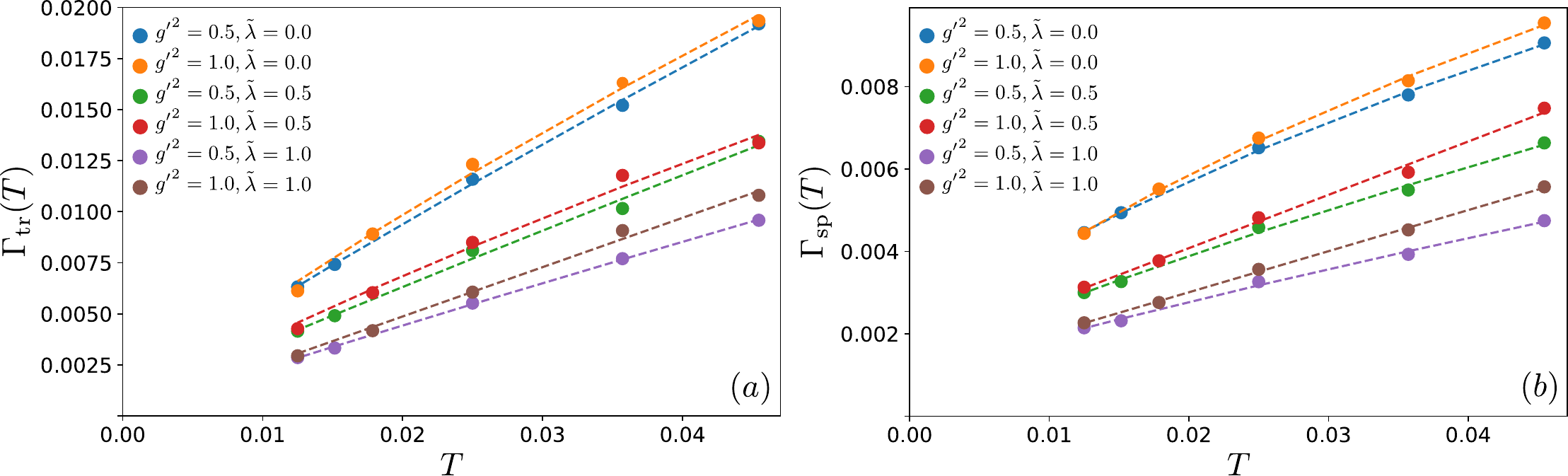}
    \caption{(a) The DC transport scattering rate $\Gamma_{\text{tr}}$ in the strange metal phase for different values of $g'$. For the same value of $\tilde{\lambda}$, the scattering rates for different values of $g'$ are nearly identical. (b) The equivalent analysis for the single particle scattering rate $\Gamma_{\text{sp}}(T)$, with the same conclusions as in (a).}
    \label{fig:Gamma_universality}
\end{figure*}

At $g' = 1$ the model shows the same features as at $g' = 1/\sqrt{2}$. A strange metal with a $T$-linear $\Gamma_{\text{tr}}(T)$ onsets at $\lambda = \lambda_G$, where the bosonic sector becomes gapless (see Appendix \ref{app:more_plots} for the evolution of the boson density of states $\overline{\nu(e)}$). It extends up to $\lambda = \lambda_s$, beyond which the $T$-dependence of $\Gamma_{\text{tr}}(T)$ becomes logarithmically corrected and a residual scattering rate appears due to order parameter freezing and the formation of SRO (Appendix \ref{app:more_plots}). The numerical values of $\lambda_s = 12.5$ and $\lambda_G = 13.5$ are different from their respective values of $5.0$ and $5.5$ at $g' = 1/\sqrt{2}$. However, for both the values of $g'$, $\lambda_s$ and $\lambda_G$ correspond to the coupling-independent values of $\tilde{\lambda} = 0$ and $\tilde{\lambda} = 1.0$ respectively. Fig. \ref{fig:Gamma_universality} shows the universality of both $\Gamma_{\text{tr}}(T)$ and $\Gamma_{\text{sp}}(T)$ in the strange metal phase: the scattering rates are nearly independent of the coupling $g'$ for the same values of $\tilde{\lambda}$. The strange metal phase has $\Gamma_{\text{tr}}(T) \approx \alpha_0(\tilde{\lambda})T$, with $\alpha_0(\tilde{\lambda} = 0) \approx 0.4$ and $\alpha_0(\tilde{\lambda} = 1.0) \approx 0.2$ at its two ends. The universality of $\Gamma_{\text{tr}}(T)$ and $\Gamma_{\text{sp}}(T)$ also holds for values of $\lambda \lesssim \lambda_s$ (shown in Appendix \ref{app:more_plots}).

\begin{figure*}
\centering
    \includegraphics[width=0.98\textwidth]{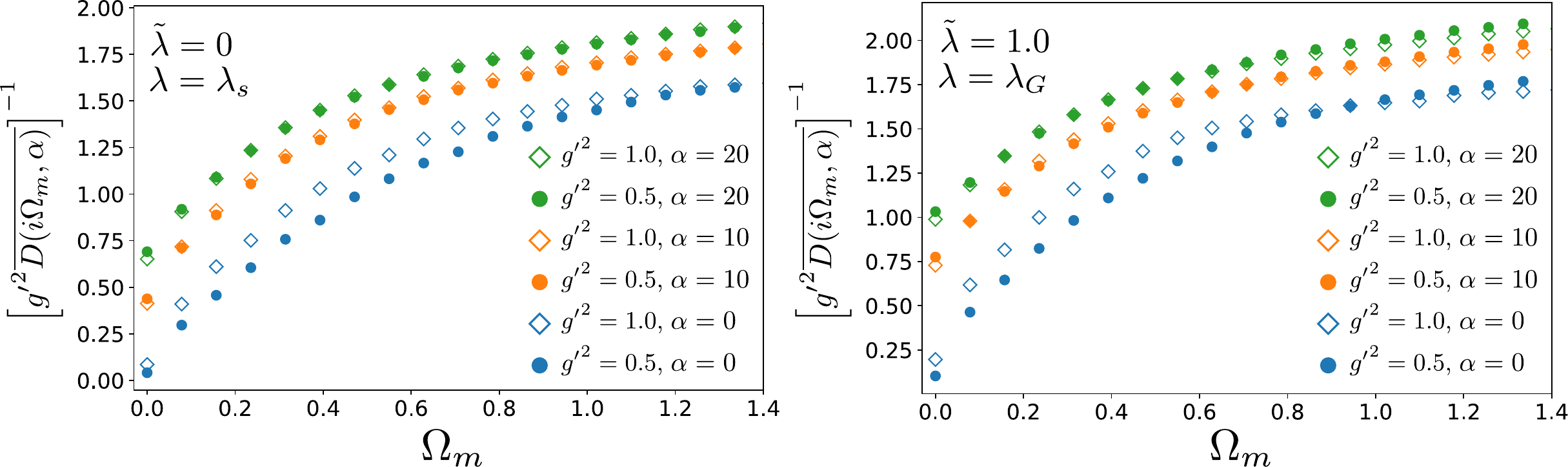}
    \caption{The inverse disorder-averaged boson propagator in the eigenbasis rescaled by the coupling ({\it i.e.} $[{g'}^2\overline{D(i\Omega_m, \a)}]^{-1}$), for different values of ${g'}$ and $\tilde{\lambda}$. For a given value of $\tilde{\lambda}$, this quantity is roughly independent of $g'$, thereby establishing that $\overline{D(i\Omega_m, \alpha)} \sim 1/{g'}^2$.}
    \label{fig:boson_prop_universality}
\end{figure*}

A rationale for the coupling-independence of the scattering rates for physically corresponding values of $\lambda$ at different $g'$ can be gleaned from the scaling of the boson propagator $\overline{D(i\Omega_m, \alpha)}$ with $g'$. As shown in Fig. \ref{fig:boson_prop_universality}, for fixed $\tilde{\lambda}$, $\overline{D(i\Omega_m, \alpha)} \sim 1/{g'}^2$ at low energies (this also leads to $\overline{\nu(e)} \sim (E_F/{g'}^2)\tilde{\nu}(e E_F/{g'}^2)$ for fixed $\tilde{\lambda}$, as shown in Appendix \ref{app:more_plots}). Since Feynman diagrams describing fermion correlation functions contain a factor that scales as ${g'}^2$ for every boson propagator, the low-energy fermion correlation functions that determine the scattering rates should also therefore be coupling-independent. These observations point towards the emergence of universal physics in the infrared that is independent of the bare coupling constants, which is a highly non-trivial feature to obtain in a disordered model, let alone one containing fermions at a nonzero density.

Earlier works \cite{Murthy2023, HardyPatelalpha0_from_Hubbard_U} have noted that planckian scattering rates in metals can occur when they are on the verge of a destructive insulating transition: when expressed as a function of a parameter that tunes the transition, the slope of the $T$-linear scattering rate reaches a maximum $O(1)$ value just before the transition occurs (which is termed a ``stability bound" on the slope). The maximal planckian coefficient $\alpha_0 \approx 0.4$ in our work occurs at $\tilde{\lambda} = 0$ ({\it i.e.} $\lambda = \lambda_s$), just before the onset of order parameter freezing. While this freezing does not destroy the metal, it does change the nature of the metallic state. Therefore, our results provide a non-trivial example of the generalization of the principles of Ref. \cite{Murthy2023} to non-destructive transitions in strongly correlated metals.

\section{Methods}
\label{sec:hmc}

This section describes various details of our computational methods. Readers who are solely interested in the physics of the problem may safely skip this section.

We use the HMC method of Ref. \cite{Lunts2023HMC} to efficiently simulate the action of Eq. (\ref{eq:action:g'}). Ref. \cite{Lunts2023HMC} simulated the same theory in the clean limit, and we refer the reader to the Methods section of that paper for details on the workings of the algorithm. Here, we detail the slight difference of the discretized actions, and the technical improvements over Ref. \cite{Lunts2023HMC} that help us reach the required low temperatures and average over several disorder realizations, while maintaining a large spatial system size.  

Once the fermion fields are integrated out, the partition function for the model with a fixed disorder configuration $g'_{\boldsymbol{r}=(x,y)}$ drawn from Eq. (\ref{eq:disorder:dist}) is given by
\begin{equation}
    \mathcal Z(g'_{(x,y)}) = \int \mathcal D \bphi \, e^{-\mathcal S_B (\bphi)} \left[\operatorname{det} A(\bphi; g'_{(x,y)})\right]^{N_f = 2}.
    \label{eq:Zd}
\end{equation}
Here, $\mathcal S_B (\bphi)$ are the terms of the action in Eq. (\ref{eq:action:g'}) containing only the bosonic $(\bphi)$ field. The fermion matrix of a single fermion flavor, $A(\bphi; g'_{(x,y)})$, appears with a power of the number of fermion flavors $N_f = 2$. The determinant $\operatorname{det} A(\bphi; g'_{(x,y)}) \geq 0$ owing to the sign-problem-free nature of the model. The partition function is therefore an integral over a legitimate positive-definite probability density, that is sampled via HMC. Observables are then computed by averaging over the $\bphi$ field first, followed by an average over disorder, as in Eq. (\ref{eq:disorder:avg}). 

The fermion matrix $A(\boldsymbol \phi; g'_{(x,y)})$ is a function of both the dynamical $\boldsymbol \phi$ field and the quenched disorder field $g'_{(x,y)}$. It is written as
\begin{align}
\label{eq:D_formula}
    & A(\boldsymbol \phi; g'_{(x,y)})_{(\a, \s, x, y, \tau), (\a', \s', x',y',\tau')} = 
    \nonumber \\ &
    \delta_{\a,\a'} \delta_{\s,\s'} 
    \Big\{
    \delta_{x,x'} \delta_{y,y'} 
    \Big[\frac{1}{2} \delta_{\tau+1,\tau'} (1 - 2 \, \delta_{\tau,N_{\tau} - 1})
    \\ &
    -\frac{1}{2} \delta_{\tau-1,\tau'} (1 - 2 \, \delta_{\tau,0}) -\delta_{\tau,\tau'}(\Delta \tau \, \mu_{\a}) 
    \Big]
    \nonumber \\ &
    - \delta_{\tau,\tau'} \, \Delta \tau \, t_{\a,(x,y),(x',y')} \Big\}
    \nonumber \\ &
    + \delta_{x,x'} \delta_{y,y'} \delta_{\tau,\tau'} \; \Delta \tau \; g'_{(x,y)} \; e^{i \boldsymbol{Q}_{AF} \cdot (x,y)} \; \boldsymbol{\phi}_{\tau, (x, y)} \cdot \boldsymbol{\tau}_{\s, \s'} \, \sigma^{x}_{\a, \a'}. \nonumber
\end{align}
Here, the indices $(\a,\s,x,y,\tau)$ correspond to band, spin, $x$-coordinate, $y$-coordinate, and discretized imaginary time slice respectively, with $\sigma^x$ referring to the Pauli-$x$ matrix. The imaginary time axis is discretized into $N_{\tau}$ slices with a Trotter step size of $\Delta \tau$, with $\beta = N_{\tau} \Delta \tau$. The $x,~y,~\tau$ coordinates are defined modulo $L,~L,~N_\tau$ respectively, thereby implementing periodic boundary conditions. The only differences from the fermion matrix of Ref. \cite{Lunts2023HMC} (where it is called $D$ instead of $A$) are that (i) the discretized time derivative is now symmetric and (ii) the space-independent coupling $g$ is replaced by $g'_{(x,y)}$.

For completeness, we also provide the discretization of the purely bosonic part of the action in Eq. (\ref{eq:action:g'}) on the lattice:
\begin{align}
&\mathcal{S}_{\boldsymbol{\phi}}^{\mathrm{lat}} = \frac{\Delta \tau}{2} \sum_{x = 0}^{L - 1}\sum_{y = 0}^{L - 1}\sum_{\tau = 0}^{N_\tau - 1}\Bigg[\left(4 + \frac{2}{c^2\Delta\tau^2} + \lambda\right)|\boldsymbol{\phi}_{\tau, (x, y)}|^2 - \nonumber \\
&\boldsymbol{\phi}_{\tau, (x,y)}\cdot\Bigg(\boldsymbol{\phi}_{\tau, (x+1,y)} + \boldsymbol{\phi}_{\tau, (x-1,y)} + \boldsymbol{\phi}_{\tau, (x,y+1)} + \boldsymbol{\phi}_{\tau, (x,y-1)} \nonumber \\ 
&+\frac{\boldsymbol{\phi}_{\tau + 1, (x,y)}  + \boldsymbol{\phi}_{\tau - 1, (x,y)}}{c^2\Delta\tau^2}\Bigg)  + \frac{u}{2}|\boldsymbol{\phi}_{\tau, (x, y)}|^4 \Bigg],     
\end{align}
where periodic boundary conditions are again applied. We use $\Delta\tau = 0.1$; in Appendix \ref{app:more_plots} we demonstrate that universal aspects of the low-energy physics are not affected by the value of $\Delta\tau$ as long as it is small enough.

In the HMC algorithm, the fermion determinant is re-exponentiated using bosonic fields $\varphi, \varphi^\ast$ called `pseudo-fermions',
\begin{equation}
\left[\operatorname{det} A\right]^{N_f} = \int \mathcal{D}\varphi\mathcal{D}\varphi^\ast e^{-\varphi^\ast \left[(AA^\dagger)^{N_f/2}\right]^{-1} \varphi},    
\end{equation}
thereby trading the calculation of the determinant for the calculation of the inverse of $(AA^{\dagger})^{N_f/2}$. Importantly, this is why we require $N_f$ to be even, as we do not currently have the capability to efficiently calculate fractional powers of matrices. The computationally expensive step of the algorithm then becomes the solution of the linear system
\begin{equation}
AA^\dagger \eta = \varphi.
\label{eq:linsys}
\end{equation}
Since $AA^\dagger$ is a Hermitian positive definite matrix, we solve this linear system via the iterative conjugate gradient (CG) method.

The methodological developments of Ref. \cite{Lunts2023HMC} mainly focused on reducing the \textit{statistical} cost of running the HMC algorithm, implemented by learning optimal hyperparameters from the data in the warmup phase of the Monte Carlo runs, and therefore reducing the total number of times Eq. (\ref{eq:linsys}) needs to be solved. In this work, we focus on reducing the \textit{algorithmic} cost, i.e. the computational cost of a single solve of Eq. (\ref{eq:linsys}), by accelerating the CG algorithm. This is done in various ways that are explained as follows:  

\subsection{Preconditioned matrix inverse}

The number of iterations required to solve Eq. (\ref{eq:linsys}) for $\eta$ to some fixed tolerance via the CG method grows with the condition number $\kappa$ of $AA^\dagger$ as $O(\sqrt{\kappa})$. In a strongly interacting and gapless fermion system, such as that considered in this work, $\kappa$ grows rapidly as the temperature is reduced, eventually making the CG untenable as $T\rightarrow 0$. Therefore, a means to reduce the effective value of $\kappa$ is required in order to probe physics at very low temperatures. A common strategy to reduce the number of CG iterations needed to converge to a solution is to precondition $AA^\dagger$ with a linear operator $M$, that is both close to the inverse of $AA^\dagger$ and easy to evaluate. A good preconditioner has the property that the condition number of $M AA^{\dagger}$ is significantly less than that of $AA^\dagger$. 

For the non-disordered version of our problem with a translationally invariant Yukawa coupling that is not too large, the simplest preconditioner $M = (A_0 A_0^{\dag})^{-1}$, where $A_0$ is obtained by setting $g = g' = 0$ in Eq. (\ref{eq:D_formula}), proved effective \cite{Lunts2023HMC}. However, in the case of spatially random interactions, this is no longer the case as the free fermion matrix $A_0$ is a particularly poor approximation of the full $A$ due to the destruction of quasiparticles around the entire FS (as opposed to only near the ``hot spots" in the non-disordered case \cite{Metlitski2010}). We must therefore use more advanced preconditioners. 

We find that in practice it works quite well to take a weighted Jacobi preconditioner (which is also used in the initial smoothing steps of more optimal multigrid preconditioners that will be discussed elsewhere \cite{PatelAGM}). In this case, we can define the preconditioner as
\begin{equation}
\begin{split}
    M &= w \operatorname{diag}({AA^{\dagger}})^{-1} 
    \\ &
    \sum_{k=0}^{n-1}(I - w \operatorname{diag}({AA^{\dagger}})^{-1}AA^{\dagger})^{k} 
\end{split}
\end{equation}
where $w$ is a weighting factor (chosen to be $0.75$), $I$ is the identity matrix, and $n$ is the number of iterates of the Jacobi smoothing (usually chosen to be $n < 10$). 

\subsection{Mixed-precision CG}

The CG method can benefit from performing initial iterations with lower-precision arithmetic, and gradually increasing the precision of the arithmetic as the iterations progress. This is due to the fact that the solutions from initial iterations are poor approximations to the final solution, thus making it unneccessary to obtain them to high precision. In particular, as our linear system is solved on GPUs, the use of lower-precision arithmetic provides an especially significant speedup due to the reduced memory traffic that it entails.

In practice, we use the simplest version of such a mixed-precision solver, with one full CG solve being performed with single-precision arithmetic to obtain a solution to a relative tolerance of $10^{-5}$, with that solution used as an initial guess for a subsequent full CG solve being performed at double-precision arithmetic to the desired tolerance. 

Lower-precision approaches can also be implemented in the application of the preconditioner, as the inversion of $AA^\dagger$ provided by the preconditioner is crude and does not benefit from higher precision. We therefore run the preconditioner at half-precision in order to avoid unnecessary computation and maximize performance.

\subsection{Matrix-free methods}

\begin{figure*}
    \includegraphics[width=0.98\textwidth]{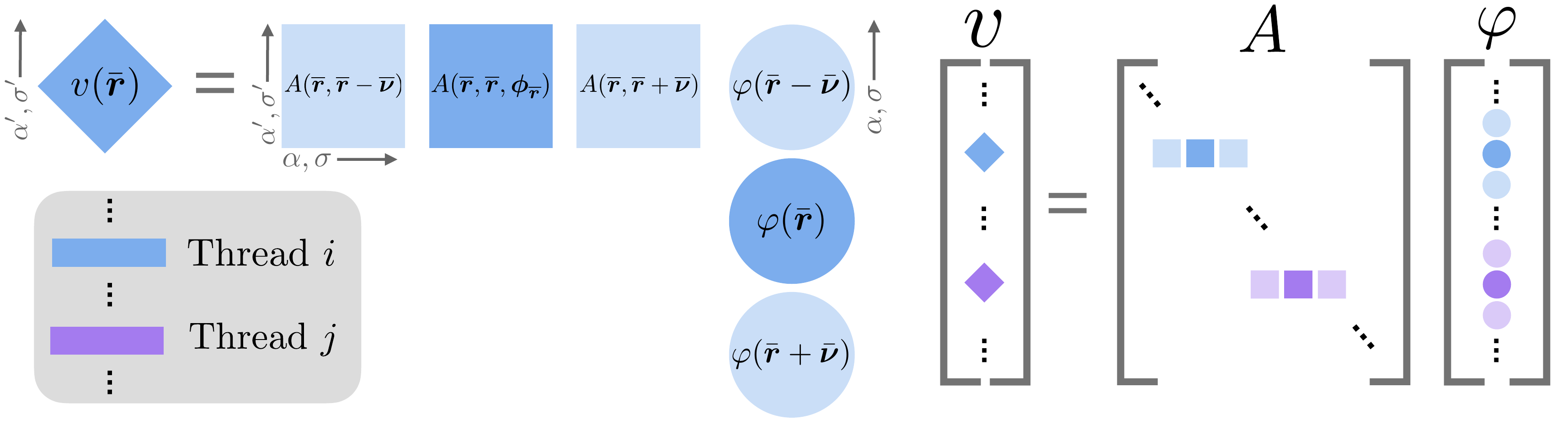}
    \caption{Schematic of the matrix-free fermion matrix-vector multiplication $v = A\varphi$. Here $\overline{\boldsymbol{r}} = (\tau, \boldsymbol{r})$ and $\overline{\boldsymbol{\nu}}$ denotes the various spacetime hopping displacements on the lattice. A \texttt{CUDA}
    kernel implementation allows for efficient parallelization of this operation on GPUs by assigning rows of the fermion matrix $A$  with maximal overlap to neighboring threads in order to exploit local caching and minimize redundant reads from the main memory. As the computation is limited by memory bandwidth, all band ($\alpha$) and spin ($\sigma$) values for a given $\overline{\boldsymbol{r}}$ are assigned to the same thread in order to eliminate redundant reads of $g'_{(x,y)} \; e^{i \boldsymbol{Q}_{AF} \cdot (x,y)} \; \boldsymbol{\phi}(\tau, x, y)$ from the main memory and thereby improve arithmetic intensity.}
    \label{fig:matvec}
\end{figure*}

The mathematical operation with the highest cost that is repeated many times in the iterative CG method is the matrix-vector multiplication of $A$ or $A^{\dagger}$ with a vector $\varphi$. Speeding up this operation is therefore important for accelerating the CG. The fastest way to do this on GPUs is to use matrix-free representations \cite{AnsorgeCUDABook} of the sparse $A$ and $A^{\dagger}$ matrices, which have a regular structure that includes a spacetime stencil. Therefore, only the factor of $g'_{(x,y)} \; e^{i \boldsymbol{Q}_{AF} \cdot (x,y)} \; \boldsymbol{\phi}(\tau, x, y)$ needs to be stored explicitly in memory in order to represent the matrix. This is illustrated schematically in Fig. \ref{fig:matvec}.

\section{Discussion}
\label{sec:discussion}

In this work we initiate a program to reproduce key experimental features of strange metal behavior in a transparent and interpretable manner, by exactly solving realistic minimal models of quantum phase transitions of metals in the presense of quenched disorder. We solve a model of a near-critical antiferromagnetic bosonic modes interacting with a FS via a quenched random Yukawa coupling, using numerically exact high-performance large-scale hybrid Monte Carlo. 

Our model exhibits an entire gapless ``quantum Griffiths phase". Within this phase, localized bosonic modes corresponding to small fluctuating antiferromagnetic puddles form, which then eventually freeze as the value of the tuning parameter $\lambda$ is reduced below $\lambda = \lambda_s$. Despite the puddles freezing, the overall spatial antiferromagnetic order remains only short-range in the entire numerically accessible parameter window, down to the lowest temperatures that we probe. 

In an extended range of $\lambda$ between the points of puddle formation and puddle ordering, the fermions form a \textit{strange metal phase}: the DC transport scattering rate and resistivity scale linearly with temperature down to our lowest temperatures. This behavior, and even the slope of the $T$-linear transport scattering rate, is independent of the strength of disordered interactions: the fermions therefore display \textit{planckian transport}. In the frequency dependence of both their dynamical single-particle and transport scattering rates, the fermions form a type of \textit{marginal Fermi liquid}, as seen in experiments \cite{VarmaMFL, Michon2022}. 

\subsection{Further comparisons with experimental phenomenology and proposals for future experiments}

Strange metal behavior has been observed extensively in cuprates near optimal hole and electron doping, both of which are away from the long-range AFM order setting in at smaller doping \cite{CHUholedopedCuprate,GreeneARCMP2020}. In between those two doping values, short-range AFM order is also commonly observed \cite{AndoPRL2001,ZXShenPNAS2019,DamascelliNPJqm2020}. Furthermore, extended regions of strange metal behavior have been found upon suppression of superconductivity with magnetic fields \cite{Cooper2009}. Our results reproduce all of these features (Fig. \ref{fig:phase_diag}). 

A recent experiment \cite{RamshawNature2021} extracted a momentum-dependent transport scattering rate in Nd-LSCO via angle-dependent magnetoresistance measurements. It found that all of the momentum-dependence was in the elastic scattering rate, with the inelastic planckian component remaining momentum-independent. This is fully consistent with the physics of our model, where the inelastic scattering rate is isotropic and planckian (as evidenced by the isotropic fermion self-energy in Fig. \ref{fig:Sigma_multiple_k}). In retrospect, this isotropy explains why the Drude formula with an inelastic scattering rate is able to describe transport properties, both in our model (Eq. (\ref{eq:Drude})) and in experiments (Ref. \cite{Michon2022}).

The strange metal phase of our model surprisingly hosts a very `clean' fermion sector: it has a sharp FS, large mean free path, and a near-vanishing residual resistivity. These features are also observed experimentally in several `clean' cuprates that have been prepared with relatively few structural defects in the copper-oxygen (CuO) planes \cite{MackenzieTL22011997,Hussey_2013,MackenziePRB1996}. The clean fermion sector in our model is remarkable, as the bare energy scale associated with the disorder, {\it i.e.} ${g'}^2$, is a significant fraction of the Fermi energy ($E_F/5$ to $2E_F/5$). If an equivalent level of disorder were to be inserted into the chemical potential instead, it would generate a large residual resistivity (or even localize the fermions at low enough temperatures). Therefore, the interaction disorder $g’$ clearly affects fermion dynamics and transport via a very different mechanism from conventional potential disorder. The similarity between our computed scattering rates and experimentally measured ones provides further evidence that strange metals arise due to interaction disorder, and have little to do with potential disorder. 

The large amount of disorder in our model manifests itself in the bosonic sector instead of the fermionic sector, leading to localized gapless bosonic modes. Experiments should therefore search for the effects of disorder in two-fermion bosonic quantities instead of single-fermion quantities, which is where we expect the hidden effects of inhomogeneity coming from inter-layer dopants to manifest themselves strongly.

The localized boson modes in our model are puddles of local AFM fluctuations of the size of a few lattice sites. Imaging these puddles directly would solidify the connection of our theory to real materials, and would be an enormous advance in the understanding of materials hosting strange metals. Ideally, the full time and space-resolved correlation functions of these magnetic modes should be measured. This would require a two-probe experiment to measure the non-translationally-invariant correlators, which is currently a challenge for all available techniques. More realistic is the measurement of spatially local dynamical correlations, which can still provide information about the size and dynamics of localized AFM puddles. Magnetometry using nitrogen-vacancy (NV) centers in diamond allows for the imaging of the microscopic magnetic structure of materials, potentially down to the nanoscale \cite{Jayich2016NVmagnet}. More recently, a scanning tunnelling microscope (STM) with an organic spin-$1/2$ molecule attached to its metallic tip has been used to measure magnetism down to the atomic scale \cite{Esat_MagSTM2024}. These methods therefore provide a promising way to detect the aforemetioned AFM puddles. 

We also note that inhomogeneous puddles of other correlation-induced orders have already been observed in cuprates: (i) puddles of charge density wave (CDW) order have been observed using high-resolution X-ray diffraction imaging \cite{Ricci_CDW_puddles_2015} and (ii) superconducting puddles have been observed using STM \cite{Allan_SC_puddles_2023}. The existence of these puddles lends credence to the possibility of similar puddles of AFM order. Furthermore, if the aforementioned CDW puddles are coupled to AFM (such as in the case of stripe order), electron energy loss spectroscopy (EELS) that is performed both locally \cite{goodge2020atomicresEELS} and in a spin-resolved manner \cite{Kuwahara_spinresEELS2011} can be used to detect the magnetic components. 

The inhomogeneous puddles will likely also lead to inhomogeneities in the transport scattering rate, causing its local values to be different from its average value of $\Gamma_{\text{tr}}$. This will lead to inhomogeneous Joule heating, which can be imaged using thermal imaging techniques such as those recently employed for graphene \cite{OkazakiLocalHeat}, serving as a direct probe of the transport consequences of the puddles.

Finally, an interesting feature of our model is the freezing of the localized short-range AFM fluctuations into glassy AFM SRO, which onsets exactly at the end of the strange metal phase where the strange metal scattering rate is the largest. Such freezing is also observed in LSCO, where the quasi-static AFM SRO (which is also glassy in nature) was found to begin exactly at the value of doping associated with the onset of the pseudogap at the end of the strange metal phase \cite{Frachet_SpinGlassPseudogapNatPhys2020, campbell2024strangemetalspinfluctuations}, which in turn is the doping value at which the strange metal behavior is most prominent. Additionally, a rapid rise in the residual resistivity as the SRO takes hold occurs both in our model as well as in experiments on LSCO \cite{Tranquada2024LSCO}.

\subsection{Future directions in modeling}

In this work we have left out various possible terms in our action, in order to clearly discern the effects of the disordered Yukawa coupling $g'_{\boldsymbol{r}}$ on the physical observables. An important next step is to (sequentially) reintroduce them into our modeling. The random chemical potential $\mu'_{\boldsymbol{r}}$ is the most commonly studied effect of quenched disorder in materials. As a moderate $\mu'_{\boldsymbol{r}}$ only contributes to elastic scattering, we expect it to mostly affect the residual resistivity and not the $T$-dependent part of $\rho(T)$. Furthermore, even if a large enough $\mu'_{\boldsymbol{r}}$ indirectly generates a contribution to $g'_{\boldsymbol{r}}$, it should not affect the universal linear-in-temperature slope of $\rho(T)$ that is independent of $g'_{\boldsymbol{r}}$ for fixed $\tilde{\lambda}$. The insensitivity/sensitivity of the linear-in-temperature slope/residual resistivity to disorder induced by electron irradiation is exactly what is observed in the cuprates \cite{Rullier-Albenque_EPL2000_radiation}, and demonstrating this effect numerically using $\mu'_{\boldsymbol{r}}$ would definitively identify a source of physical disorder with a model hamiltonian parameter. 

Another crucial parameter is the translationally invariant Yukawa coupling $g$, which in a real material would likely be in the regime $g \gg g'$. Earlier theoretical work suggests that it will not contribute to the transport scattering rate \cite{HublinaRicePRB1995}, but this has never been solidified. This two-component coupling might also provide an explanation for the observations from Ref. \cite{RamshawNature2021}, in which the inelastic/elastic part of the scattering rate is momentum-independent/momentum-dependent. Furthermore, attractive interactions arising from $g$ and $g'_{\boldsymbol{r}}$ will induce pairing in different channels (d-wave vs. s-wave), and this competition will have interesting consequences for superconductivity. Additionally, the non-trivial momentum space structure generated by the AFM ordering wavevector $\boldsymbol{Q}_{\text{AF}}$ from the part of the interaction vertex proportional to $g$ could be important for pseudogap physics. 

Similarly important is the introduction of spatial correlation in the disordered interactions, $\overline{g'_{\boldsymbol{r}_1} g'_{\boldsymbol{r}_2}} \neq \delta_{\boldsymbol{r}_1,\boldsymbol{r}_2}$. Intrinsic structural correlations are ubiquitous in doped compounds such as cuprates \cite{Greven2024structuralcorrelations, Ricci_CDW_puddles_2015}. Correlated potential disorder has also been argued to bolster superconductivity \cite{NeverovNatComm2022_correlated_disorder}. Like a nonzero $g$, correlations in $g'_{\boldsymbol{r}}$ would also reintroduce the ordering wavevector and therefore could also play a role in the physics of the pseudogap.

Ultimately, we are interested in simulating the most realistic model possible. This includes the theory without the doubling of fermion flavor, i.e. $N_f = 1$. From perturbative arguments, we expect this theory to have an effective coupling strength in the infrared that is two times larger. Naively, this would lead to a doubling of the maximal planckian coefficient to $\alpha_0 \approx 0.8$, which would be closer to the $\alpha_0 \sim \mathcal{O}(1)$ values observed in cuprates \cite{Legros2019, RamshawNature2021}. Additionally, a larger effective coupling strength should lead to an increase of the superconducting transition temperature, possibly pushing superconductivity (which is absent in our present study) into our numerically accessible temperature window. However, in order to simulate such a theory, we need the ability to compute inverse square roots of the fermion matrices. Such computations generally fall under the ambit of rational HMC \cite{clark2006rational}, and are a technical advance to be made in future work. 

An additional realistic modification would be the inclusion of a repulsive Hubbard interaction term $U$. Such a term would introduce a sign problem into the Monte Carlo simulations, rendering it beyond the scope of our numerical method. However, a recent work has used dynamical mean field theory (DMFT) to investigate a model of electrons with Hubbard $U$ repulsion coupled to a 2D bosonic bath via random Yukawa interactions that are treated in an averaged manner using the replica trick \cite{HardyPatelalpha0_from_Hubbard_U}. They find a strange metal (although only at a QCP), where the scattering rate is $\Gamma \approx \alpha_0 k_B T/\hbar$ with $\alpha_0$ strongly enhanced by a nonzero $U$. Whether or not such an enhancement would occur in our exact treatment of the random Yukawa interactions is an open question, and an exact simulation of our model in the presence of a Hubbard $U$ is therefore highly desirable. A Hubbard $U$ is likely also required in order to connect the low-$T$ strange metal in our model to the ``bad metal" physics that occurs in experiments at much higher temperatures, in which the 2D resistivity approaches (and sometimes exceeds) the resistance quantum of $\hbar/e^2$ \cite{Deng2013}.

Since our model has no bare boson dynamics and no bare boson self-interactions, it corresponds exactly to a fermion hamiltonian with two-body interactions that is obtained by integrating out the bosons. It would therefore be interesting to examine the connections between the many-body quantum wavefunctions of such a hamiltonian  and strange metallicity, by using computational methods that provide direct access to them \cite{PatelChanglaniffn}.

In a different direction, we would like to have a materials-level understanding of the emergence of the the spatial inhomogeneity in the Yukawa coupling. We expect the main contributions to $g'_{\boldsymbol{r}}$ to come from inter-layer dopants, as these have potentials that exhibit a long-range effect on the CuO planes, modifying many hopping integrals $t_{ij}$, an effect that has already been seen via \textit{ab initio} modeling \cite{HirschfeldPRB2022}. The inhomogeneity in $t_{ij}$ should then translate to a inhomogeneity in spin-exchange interactions $J_{ij}$, which would lead to $g'_{\mathbf r}$ upon decoupling the spin-spin interaction with auxiliary bosonic variables $\boldsymbol{\phi}_{\tau,\boldsymbol{r}}$ representing the collective spin modes. 

Finally, on the technical side, there are multiple extensions of our HMC method that can be implemented. One such extension is the multigrid preconditioner \cite{ClarkAGM2010}, used in HMC simulations of lattice quantum chromodynamics (QCD) to accelerate the CG method. This needs to be adapted to fermion matrices corresponding to systems with Fermi surfaces instead of the Dirac points of zero-density QCD in order to have utility in the simulation of correlated metals, and therefore requires more careful methodological development. It will be presented in a forthcoming paper \cite{PatelAGM}. Another is the use of a variant of HMC called the No-U-Turn-Sampler \cite{JMLR:v15:hoffman14a}, which has been shown to be superior to HMC in many cases and is less costly to auto-tune \cite{osti_1430202}. 

\section*{Acknowledgements}

We thank Subir Sachdev for several helpful discussions and collaboration on related work. We also thank Fakher Assaad, Antoine Georges, Berit Goodge, David Leboeuf, Zi Yang Meng, Markus M{\"u}ller, Apoorva Patel, J{\"o}rg Schmalian and John Tranquada for useful discussions. This research was supported by the U.S. National Science Foundation grant No. DMR-2245246, by the Harvard Quantum Initiative Postdoctoral Fellowship in Science and Engineering, and by the Simons Collaboration on Ultra-Quantum Matter which is a grant from the Simons Foundation (651440, S. S.). M. S. A. is supported by a Junior Fellowship from the Harvard Society of Fellows and  the ONR project under the Vannevar Bush award ``Mathematical Foundations and Scientific Applications of Machine Learning”. This research was also supported in part by grant NSF PHY-2309135 to the Kavli Institute for Theoretical Physics (KITP). The Flatiron Institute is a division of the Simons Foundation.

\bibliography{references}

\onecolumngrid

\section*{Appendices}
\appendix

\section{Boson gap and susceptibility}
\label{app:boson_gap_susc}

\begin{figure*}
    \centering
    \includegraphics[width=0.98\textwidth]{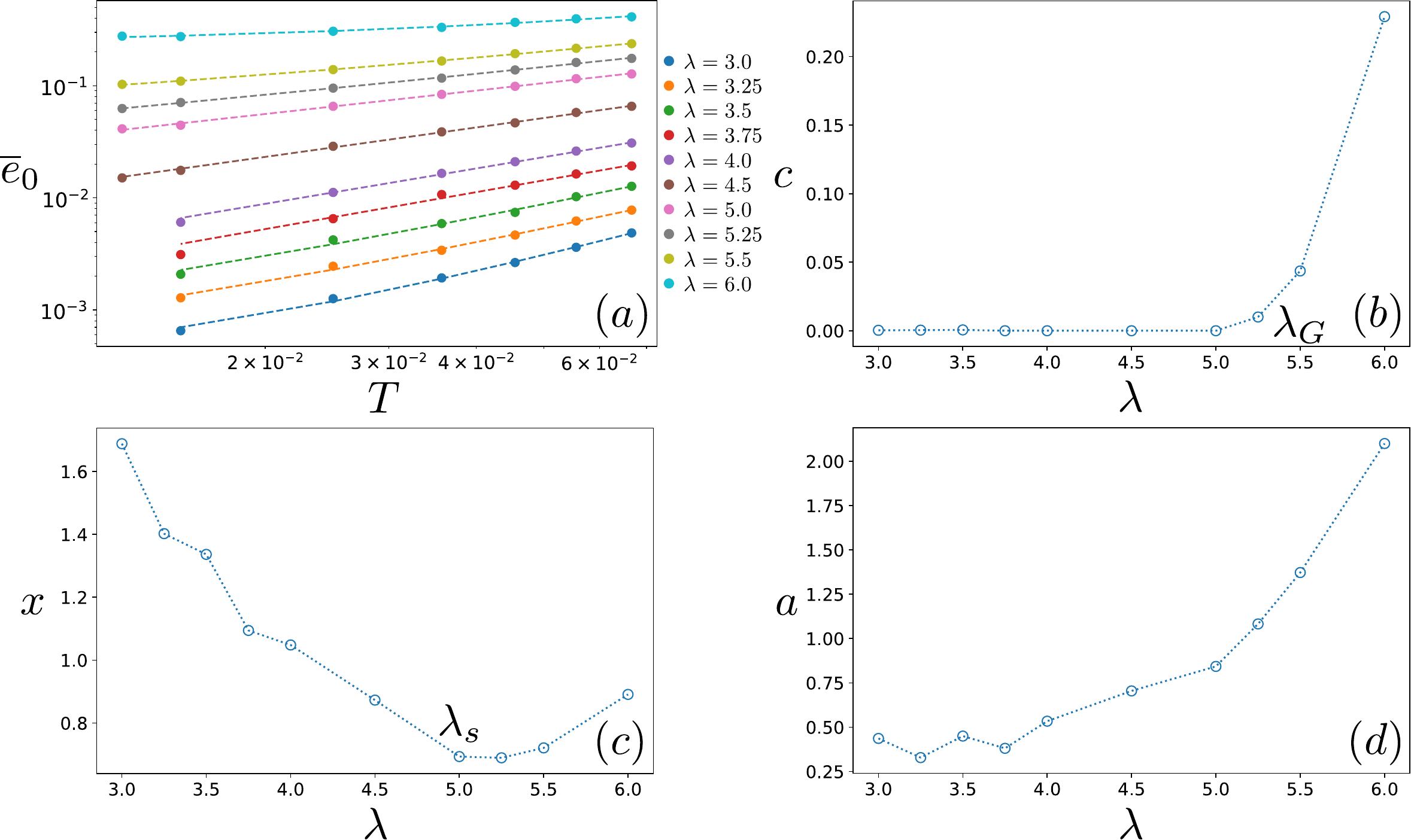}
    \caption{(a) The $T$ dependence of the disorder-averaged boson gap $\overline{e}_0$ for various values of $\lambda$. The dashed curves are a fit to $\overline{e}_0(T) = c + a T^x$. (b) - (d) The depencencies of the fit parameters on $\lambda$.}
    \label{fig:boson_gap_fit}
\end{figure*}

In this Appendix, we first analyze the $T$-dependence of the disorder-averaged boson gap $\overline{e}_0$. In a gapless phase, $\overline{e}_0(T \rightarrow 0)$ should vanish, and $\overline{e}_0(T)$ should have a power-law dependence on $T$ \cite{Sachdev_QPT}. In contrast, in a gapped phase, $\overline{e}_0(T \rightarrow 0) > 0$, and $\overline{e}_0(T)$ has a weak $T$-dependence. We therefore fit $\overline{e}_0(T)$ to $c + a T^x$ (Fig. \ref{fig:boson_gap_fit}). We find that the zero-temperature gap $c$ indeed becomes very small in the ``quantum Griffiths phase" ($\lambda < \lambda_G$). The exponent $x$ exhibits a minimum in the strange metal region $\lambda_s < \lambda < \lambda_G$, indicating that the temperature sensitivity of the gap ({\it i.e.} $d\overline{e}_0/dT|_{T \rightarrow 0}$) is the largest in this regime.  

\begin{figure*}
    \centering
    \includegraphics[width=0.98\textwidth]{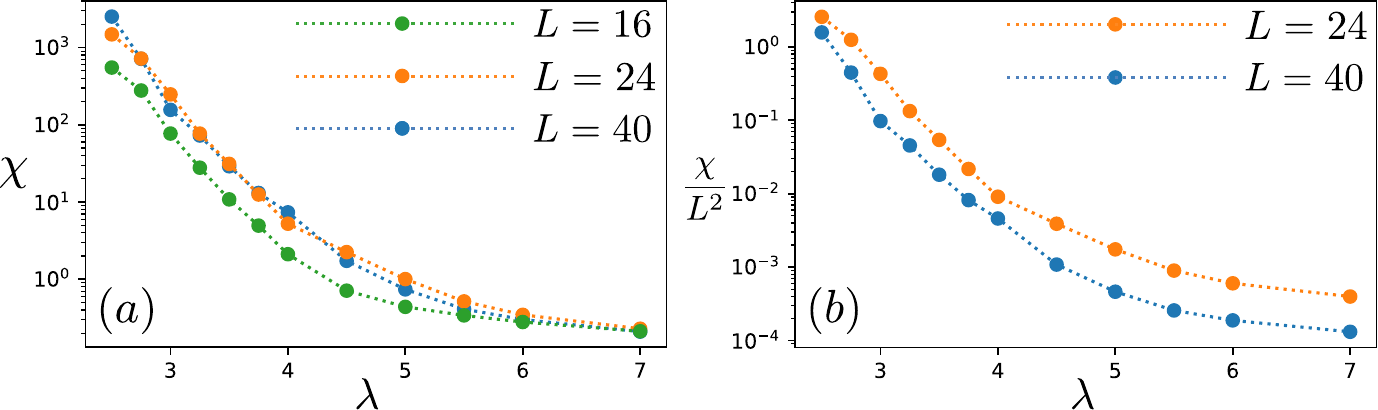}
    \caption{(a) The disorder-averaged boson susceptibility $\chi$ for various values of the system size $L$ and tuning parameter $\lambda$, at inverse temperature $\beta=66$. The value of $\chi$ increases when going from $L=16$ to $L=24$, with no further increase when going from $L=24$ to $L=40$. This indicates that finite-size effects are present at $L=16$, but become negligible for $L\geq 24$. (b) The normalized disorder-averaged susceptibility $\chi/L^2$ as a function of $\lambda$. Because $\chi$ saturates as a function of $L$ for $L\geq 24$, as shown in (a), the value of $\chi/L^2$ decreases as $L$ is increased for $L\geq 24$. Therefore, there is no extensive scaling of $\chi$ with the system volume $L^2$ for large $L$, and thus no LRO.}
    \label{fig:boson_susc_L}
\end{figure*}

Second, in order to show the absence of LRO, we analyze the disorder-averaged uniform static boson susceptibility $\chi = \overline{D(i\Omega_m = 0, \boldsymbol{q} = 0, \boldsymbol{q}' = 0)}$. We find that $\chi$ does not scale extensively in the system volume $L^2$ (Fig. \ref{fig:boson_susc_L}), implying the lack of LRO at all values of $\lambda$ and $T$ considered.

\begin{figure*}
    \centering
    \includegraphics[width=0.98\textwidth]{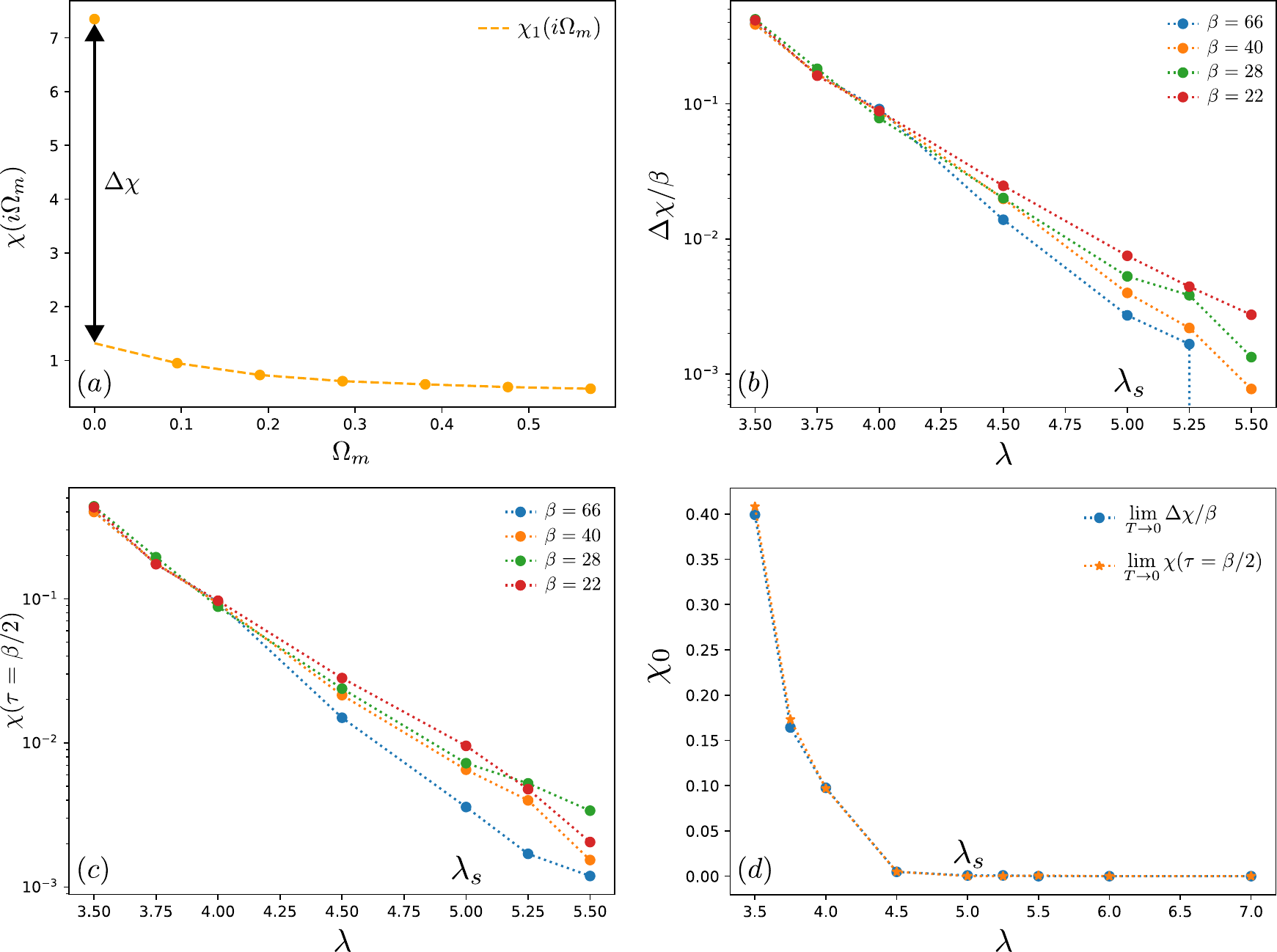}
    \caption{(a) The dynamic susceptibility $\chi(i\Omega_m)$ at $\lambda = 4.0 < \lambda_s$ and inverse temperature $\beta=66$, showing an excess susceptibility $\Delta\chi$ at $\Omega_m=0$; the value of $\Delta\chi$ is determined by taking the difference between the actual susceptibility at $\Omega_m=0$ and the value extrapolated from $\chi(i\Omega_m > 0)$ using a cubic spline. (b) $\Delta\chi/\beta$ for various $\lambda$ and $\beta$; a crossover to $\beta$-independent behavior of $\Delta\chi/\beta$ is observed for $\lambda < \lambda_s$. (c) $\chi(\tau = \beta/2)$ for various $\lambda$ and $T$, with the same behavior as in (b). (d) The frozen component $\chi_0$ of the susceptibility, which is nonzero for $\lambda < \lambda_s$, derived by (i) linearly extrapolating $\Delta\chi/\beta$ to $T=0$ and (ii) linearly extrapolating $\chi(\tau = \beta/2)$ to $T=0$; both methods produce the same value.}
    \label{fig:boson_dynamic_susc}
\end{figure*}

Third, for $\lambda < \lambda_s$, the disorder-averaged uniform dynamic susceptibility $\chi(i\Omega_m) = \overline{D(i\Omega_m, \boldsymbol{q} = 0, \boldsymbol{q}' = 0)}$ ($\chi(\tau)$ in the imaginary time domain) displays behavior corresponding to the partial freezing of the order parameter $\boldsymbol{\phi}$ in time. We see that $\chi(i\Omega_m) = \Delta\chi\delta_{\Omega_m,0} + \chi_1(i\Omega_m)$ (Fig. \ref{fig:boson_dynamic_susc} (a)), with $\Delta\chi$ crossing over to $\Delta\chi \propto \beta = 1/T$ as $\lambda$ is reduced below $\lambda_s$ (Fig. \ref{fig:boson_dynamic_susc} (b)). This leads to a nonzero $\chi(\tau=\beta/2 \rightarrow \infty)\approx \Delta\chi/\beta$ in the $T\rightarrow 0$ limit for $\lambda < \lambda_s$ (Fig. \ref{fig:boson_dynamic_susc} (d)), which is the frozen ({\it i.e.} time-independent) component $\chi_0$ of the susceptibility arising from the frozen component of $\boldsymbol{\phi}$. The frozen $\boldsymbol{\phi}$ component only produces SRO and not LRO because of the short localization length of the low-lying bosonic eigenmodes (see Fig. \ref{fig:LLs_and_eigenmodes}).

\begin{figure*}
    \centering
    \includegraphics[width=0.98\textwidth]{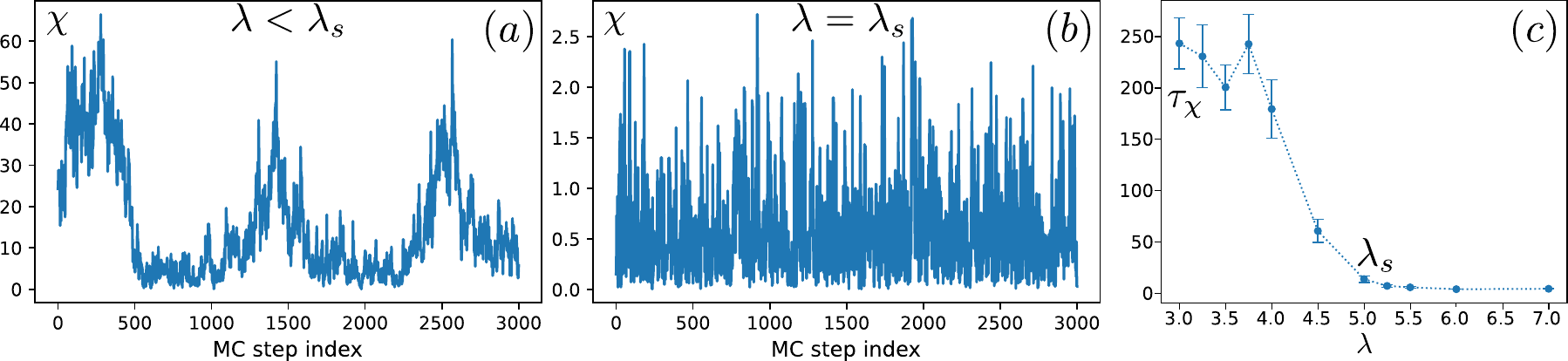}
    \caption{A typical Monte Carlo (MC) time series of the susceptibility $\chi$ for (a) $\lambda = \lambda_\ast < \lambda_s$ and (b) $\lambda = \lambda_s$. (c) The integrated autocorrelation time for the susceptibility time series, $\tau_\chi$, as a function of $\lambda$, showing a sharp increase as glassiness sets in for $\lambda < \lambda_s$.}
    \label{fig:chi_MC_dynamics}
\end{figure*}

Lastly, in order to further understand the nature of the SRO for $\lambda < \lambda_s$, we look at the Monte Carlo dynamics of the susceptibility $\chi$ as a function of $\lambda$. As shown in Fig. \ref{fig:chi_MC_dynamics} (a), the susceptibility displays slow dynamics with metastable behavior as a function of Monte Carlo time for $\lambda < \lambda_s$, in contrast to the fast uncorrelated dynamics seen for $\lambda \geq \lambda_s$ in Fig. \ref{fig:chi_MC_dynamics} (b). This also leads to a large increase in the integrated autocorrelation time for $\lambda < \lambda_s$ (Fig. \ref{fig:chi_MC_dynamics} (c)). All these features are signatures of glassiness of the AFM SRO, which may be motivated as follows: integrating out the fermions results in long-range boson couplings of forms such as $g'_{\boldsymbol{r}}g'_{\boldsymbol{r}'}e^{i\boldsymbol{Q}_{\mathrm{AF}}\cdot(\boldsymbol{r} - \boldsymbol{r}')}\boldsymbol{\phi}_{\tau,\boldsymbol{r}}\cdot\boldsymbol{\Pi}(\tau, \tau', \boldsymbol{r}, \boldsymbol{r}')\cdot\boldsymbol{\phi}_{\tau',\boldsymbol{r}'}$, which fluctuate randomly in sign as a function of $\boldsymbol{r}$ and $\boldsymbol{r}'$ due to the randomess of $g'_{\boldsymbol{r}}$. Such random-in-sign long range couplings are the essence of spin glass order (see, {\it e.g.}, Ref. \cite{GeorgesGlass} and references within), and have also been proposed to produce ``cluster glass" order in mean-field calculations on disordered itinerant electron systems \cite{VladGlass1, VladGlass2}, which our results for $\lambda < \lambda_s$ appear to be a numerically exact example of.

\section{Additional data}
\label{app:more_plots}

In this Appendix, we provide additional data for various quantities discussed in the main text.

\begin{itemize}

\item In Fig. \ref{fig:eigenbasis_boson_propagator_more_lambdas}, we show $\overline{D(i\Omega_m,\a)}^{-1}$ at values of $\lambda$ on either side of $\lambda_s = 5.0$ (as done in Fig. \ref{fig:eigenbasis_boson_propagator} for $\lambda = \lambda_s$). 

\begin{figure*}
\centering
    \includegraphics[width=0.98\textwidth]{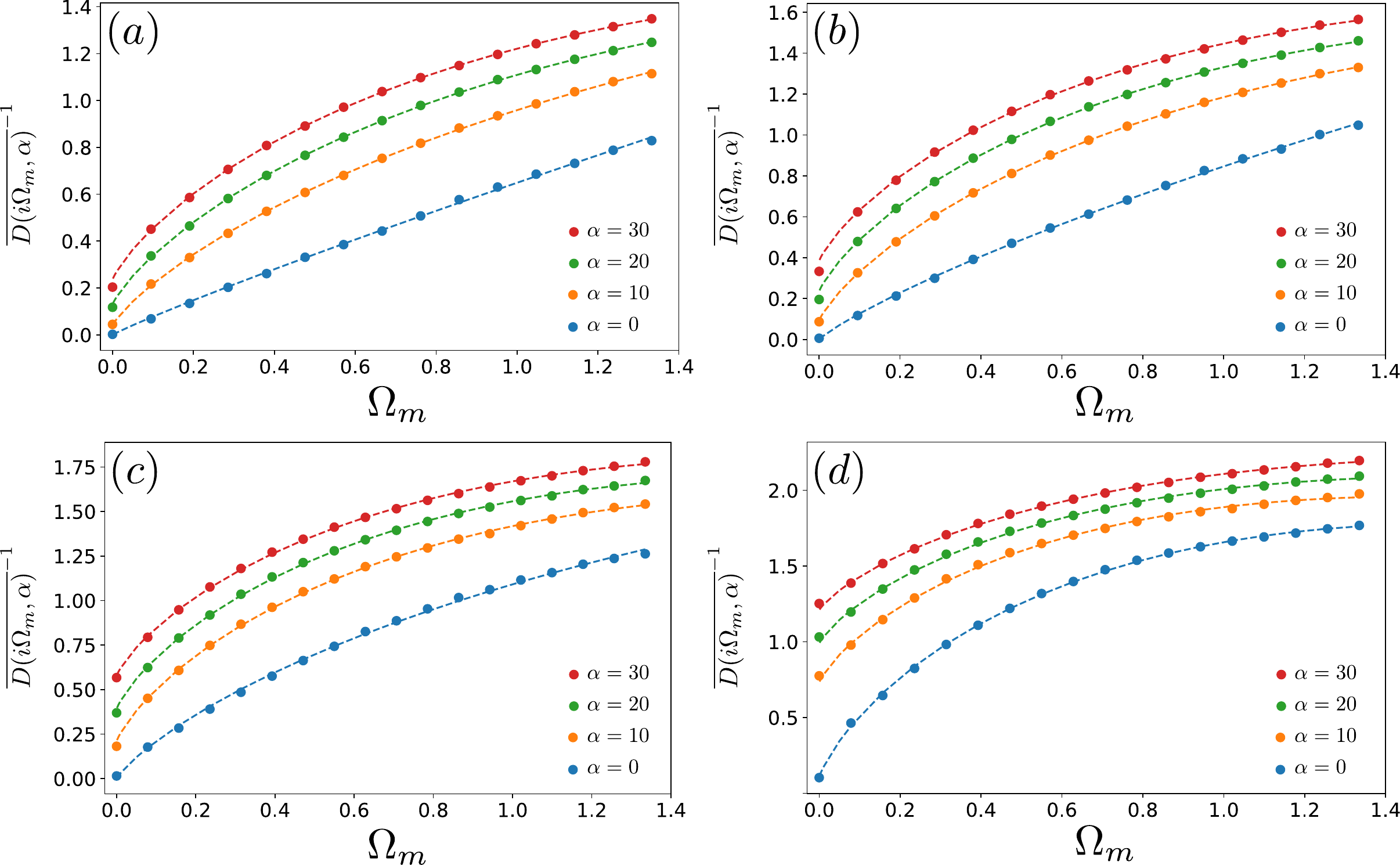}  
    \caption{The inverse disorder-averaged boson propagator in the eigenbasis, $\overline{D(i\Omega_m,\a)}^{-1}$, measured for (a), (b): $\lambda = 3.5,~4.0$ respectively at inverse temperature $\beta = 66$ and (c), (d): $\lambda = 4.5,~5.5$ respectively at $\beta = 80$. The dashed curves are fits to the scaling form discussed in the main text.}
    \label{fig:eigenbasis_boson_propagator_more_lambdas}
\end{figure*}

\item In Fig. \ref{fig:diag_boson_propagator_more_lambdas}, we show $D^{-1}(i\Omega_m, \boldsymbol{q})$ at values of $\lambda$ on either side of $\lambda_s = 5.0$ (as done in Fig. \ref{fig:diag_boson_propagator} for $\lambda = \lambda_s$). The discontinuity between the zeroth and finite Matsubara frequencies mentioned in the main text gets larger as $\lambda$ is reduced, due to increasing disorder in the boson sector at low energies as well as the formation of SRO for $\lambda < \lambda_s$. The SRO for $\lambda < \lambda_s$ is underlain by the very slow dynamics of the individual low-lying boson eigenmodes (as indicated by their small eigenvalues $e_\alpha \ll {g'}^2 T/E_F^2$ that provide relaxation timescales $\tau_\alpha \gg 1/T$) in that regime.

\begin{figure*}
\centering
    \includegraphics[width=0.98\textwidth]{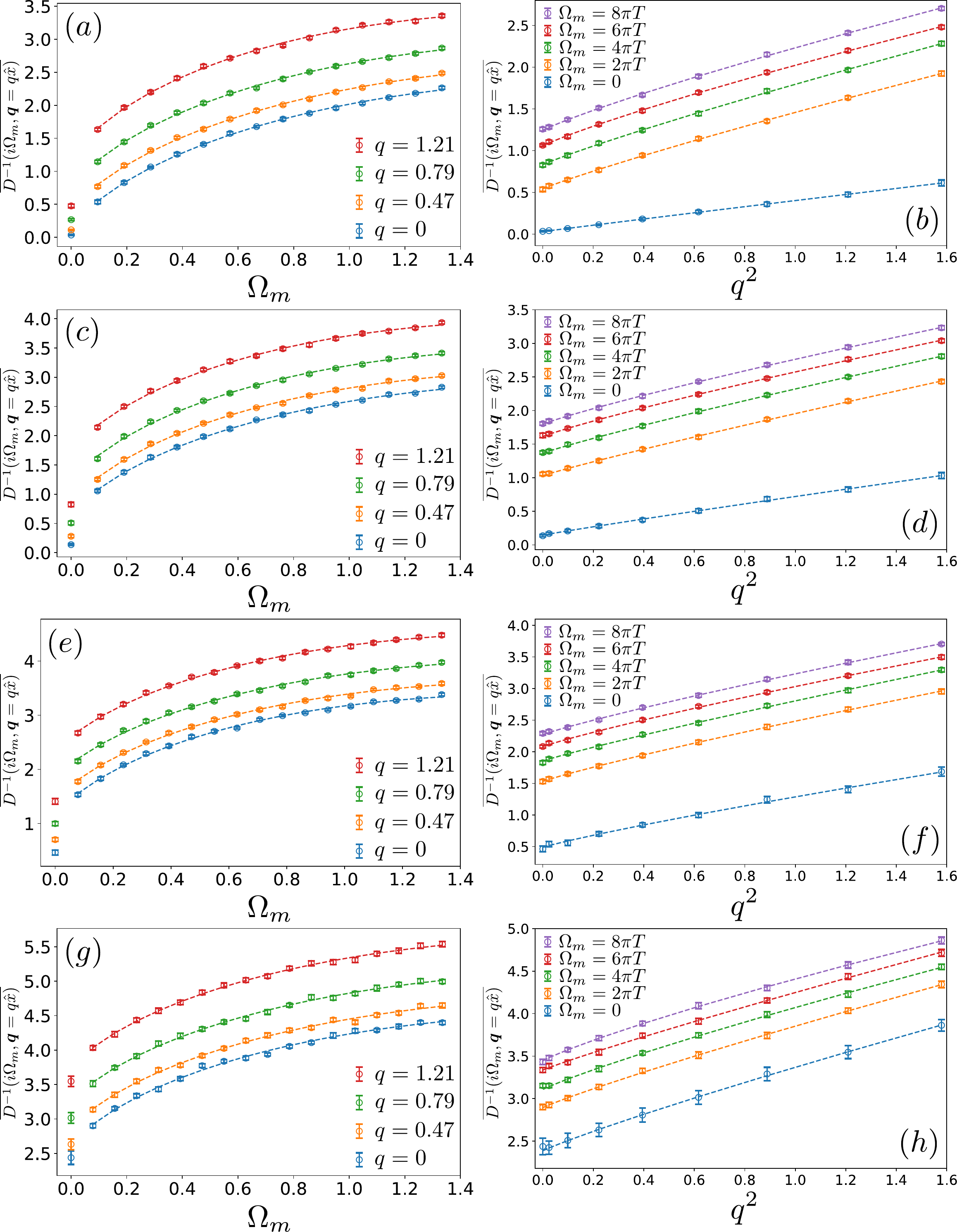}  
    \caption{The inverse disorder-averaged boson propagator in momentum space, $D^{-1}(i\Omega_m, \boldsymbol{q})$, measured for $\lambda = 3.5$ and inverse temperature $\beta = 66$ ((a), (b)), $\lambda = 4.0,~\beta = 66$ ((c), (d)), $\lambda = 4.5,~\beta = 80$ ((e), (f)), $\lambda = 5.5~\beta = 80$ ((g), (h)), and plotted as a function of $\Omega_m$ and $q^2$. The dashed curves are fits to the scaling forms discussed in the main text.}
    \label{fig:diag_boson_propagator_more_lambdas}
\end{figure*}

\item In Fig. \ref{fig:self_energy_more_lambdas} we show $-\text{Im}[\Sigma_{\text{FS}}(i\omega_n)]$ plotted vs our MFL fit Ansatz Eq. (\ref{eq:Sigma_MFL_fit}) for various values of $\lambda \leq \lambda_G = 5.5$ ({\it i.e.} in the ``quantum Griffiths phase" where the boson is gapless). The data points lie on the $y = x$ line, indicating an excellent fit.

\begin{figure*}
    \includegraphics[width=0.98\textwidth]{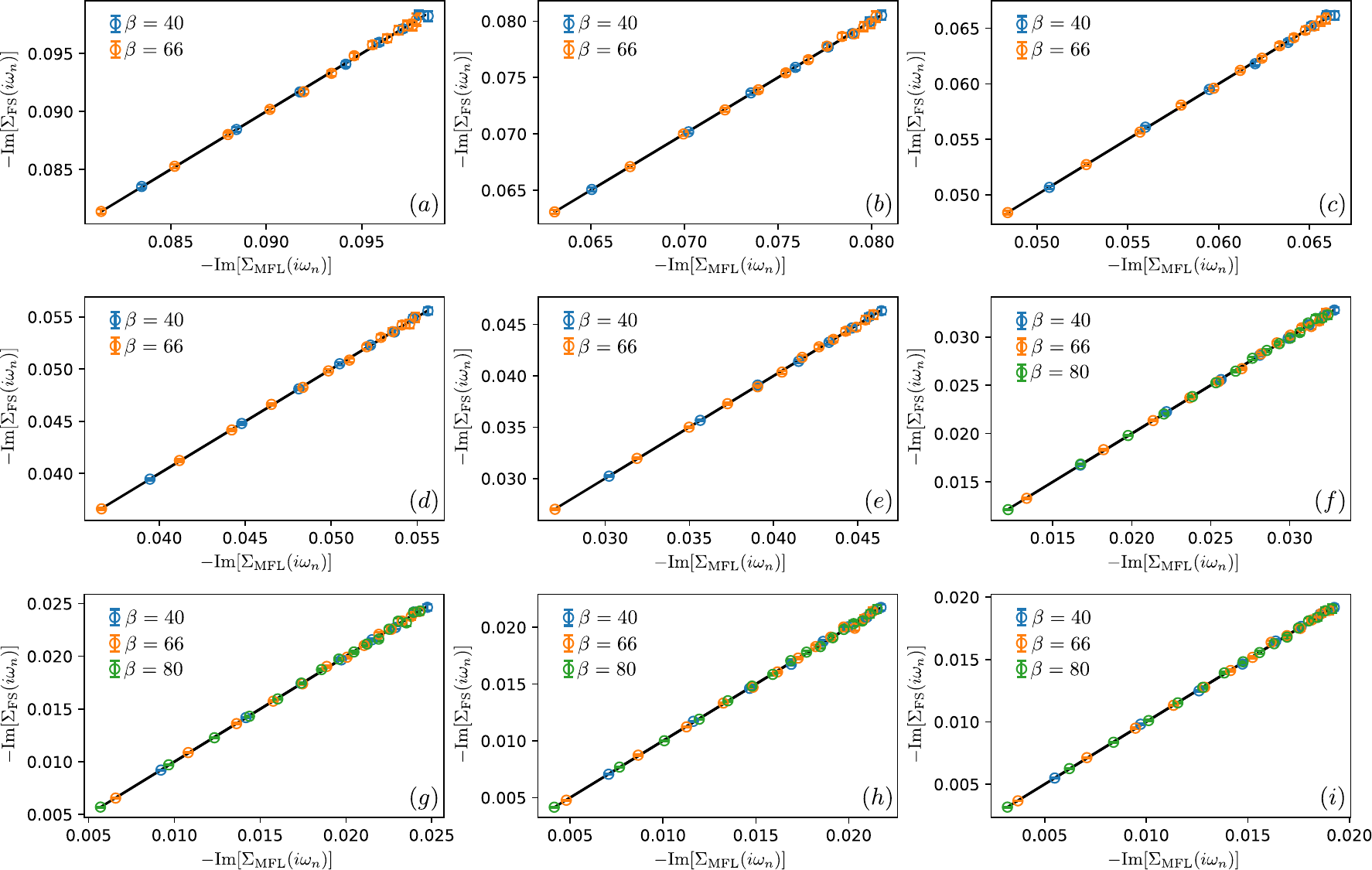}  
    \caption{The imaginary part of the Matsubara fermion self-energy averaged over the FS, $-\text{Im}[\Sigma_\text{FS}(i\omega_n)]$, at (a) - (i) $\lambda = 3.0,~3.25,~3.5,~3.75,~4.0,~4.5,~5.0,~5.25,~5.5$ respectively and different values of inverse temperature $\beta$, plotted vs the MFL fit in Eq. (\ref{eq:Sigma_MFL_fit}).}
    \label{fig:self_energy_more_lambdas}
\end{figure*}

\item Fig. \ref{fig:transport_self_energy_more_lambdas} is the equivalent of Fig. \ref{fig:self_energy_more_lambdas} for the transport self energy, $\Sigma_{\text{tr}}(i\Omega_m)$. Again, an excellent fit to the MFL ansatz (Eq. \ref{eq:Sigma_tr_MFL_fit}) is obtained over the entire ``quantum Griffiths phase".

\begin{figure*}
    \centering
    \includegraphics[width=0.98\textwidth]{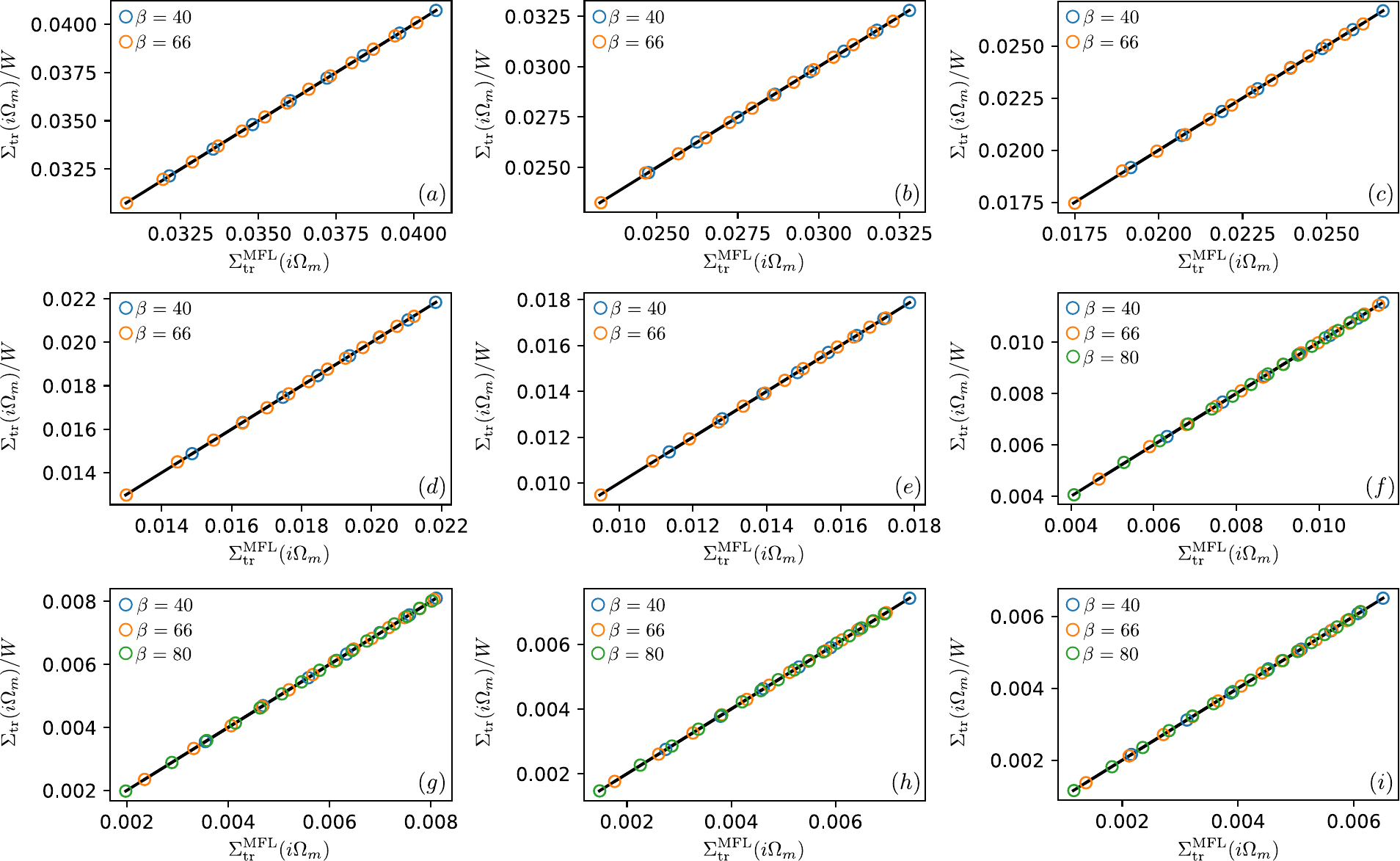}
    \caption{The ``transport self-energy" $\Sigma_\text{tr}(i\Omega_m)$ on the Matsubara frequency axis, normalized by $(1/\pi)$ times the the free Drude weight, at (a) - (i) $\lambda = 3.0,~3.25,~3.5,~3.75,~4.0,~4.5,~5.0,~5.25,~5.5$ respectively and different values of inverse temperature $\beta$ and plotted vs the MFL fit in Eq. (\ref{eq:Sigma_tr_MFL_fit}).}
    \label{fig:transport_self_energy_more_lambdas}
\end{figure*}

\item In Fig. \ref{fig:Gamma_universality_2}, we show that the universality ({\it i.e.} $g'$-independence for fixed $\tilde{\lambda}$) of the DC scattering rates also holds for $\lambda < \lambda_s$ ({\it i.e.} $\tilde{\lambda} < 0$), even though $\Gamma_{\tr}(T)$ is no longer linear-in-$T$ and the scattering rates have finite residual values in that regime.

\begin{figure*}
    \centering
    \includegraphics[width=0.98\textwidth]{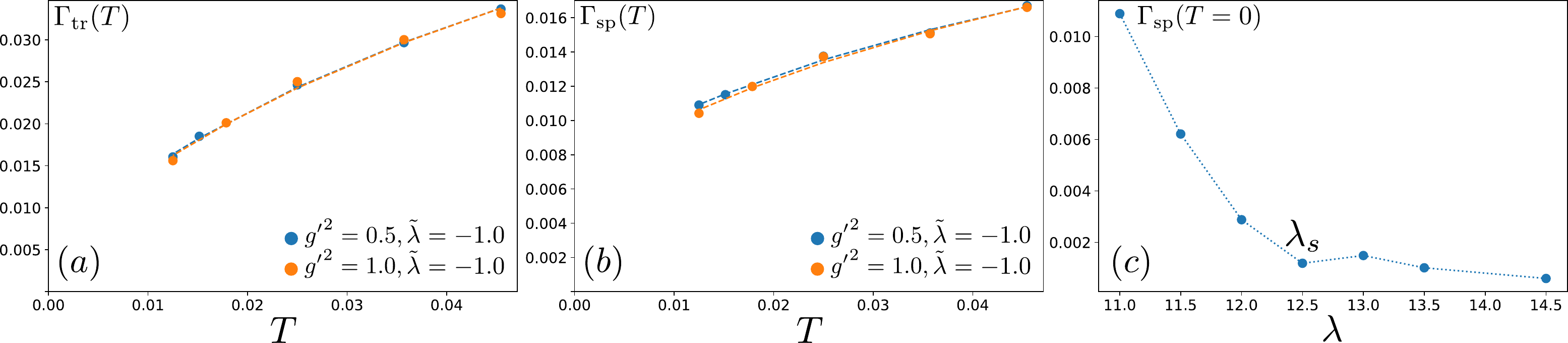}
    \caption{(a) The DC transport scattering rate $\Gamma_{\text{tr}}$ for $\lambda < \lambda_s$ and different values of $g'$. For the same value of $\tilde{\lambda}$, the scattering rates for different values of $g'$ are nearly identical. (b) The equivalent analysis for the single particle scattering rate $\Gamma_{\text{sp}}(T)$, with the same conclusions as in (a). (c) Residual single particle scattering rate $\Gamma_{\text{sp}}(T = 0)$ at $g' = 1$ as a function of $\lambda$, inferred from Eqs. (\ref{eq:Gamma_sp_FL}, \ref{eq:Gamma_sp_MFL}) in their applicable regimes.}
    \label{fig:Gamma_universality_2}
\end{figure*}

\item In Fig. \ref{fig:boson_DOS_universality}, we overlay the disorder-averaged density of states of boson eigenvalues $e$ for different values of $g'$. We find that for fixed $\tilde{\lambda}$, ${g'}^2 \overline{\nu(e)}$ is approximately a $g'$-independent function of $e/{g'}^2$, thereby establishing the scaling of $\overline{\nu(e)}$ discussed in Sec. \ref{sec:universality} in the main text. For $\lambda > \lambda_G$ ({\it i.e.} $\tilde{\lambda} > 1.0$), $\overline{\nu(e)}$ shows a clear gap, which closes at $\lambda = \lambda_G$ ({\it i.e.} $\tilde{\lambda} = 1.0$). For $\lambda < \lambda_s$ ({\it i.e.} $\tilde{\lambda} < 0$), $\overline{\nu(e)}$ shows an upturn at small $e$, which disappears at $\lambda = \lambda_s$ ({\it i.e.} $\tilde{\lambda} = 0$).

\begin{figure*}
    \centering
    \includegraphics[width=0.98\textwidth]{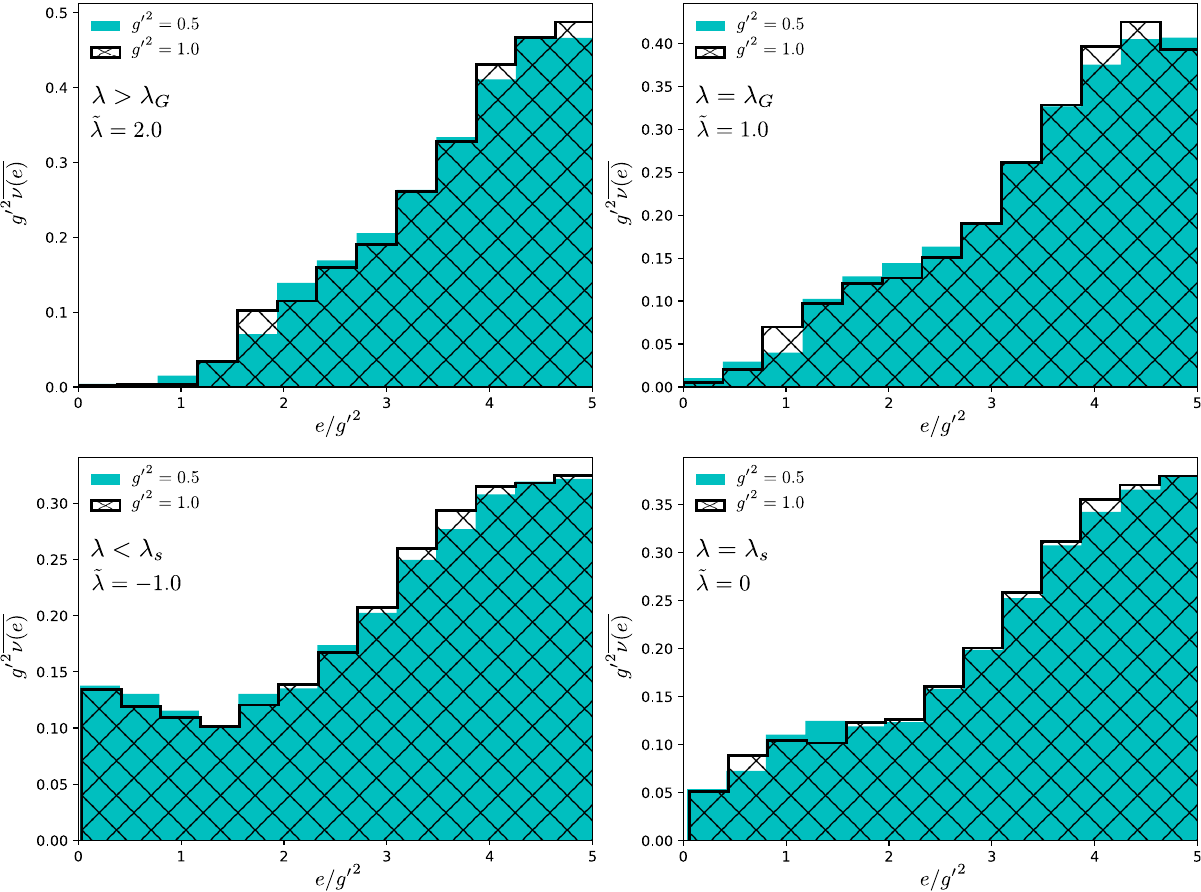}
    \caption{Rescaled disorder-averaged boson density of states, ${g'}^2 \overline{\nu(e)}$, plotted vs rescaled eigenvalues $e/{g'}^2$, for different values of $g'$ and $\tilde{\lambda}$. For a given value of $\tilde{\lambda}$, the plots are approximately independent of $g'$.}
    \label{fig:boson_DOS_universality}
\end{figure*}

\item In Fig. \ref{fig:trotter_comparison}, we demonstrate the robustness of universal low-energy physics to changes in the Trotter step size $\Delta\tau$ (which defines a UV energy cutoff of $O(1/\Delta\tau) = 10$ for the value of $\Delta\tau = 0.1$ used in this work). 

As is well known from the general theory of phase transitions and critical phenomena, universal aspects of low-energy physics such as critcal exponents and scattering rates should be independent of UV cutoffs, but non-universal quantities such as the numerical values of parameters marking the positions of QCPs can change. Moreover, since the physical UV cutoff of the Fermi energy $E_F = 2.5$ is much smaller than $1/\Delta\tau$, even the changes in non-universal quantities due to changes in $\Delta\tau$ should be fairly small, as they will be controlled by the smaller of the UV cutoffs. 

We find that this is indeed the case: changing $\Delta\tau$ from $0.1$ to $0.05$ (and therefore the UV cutoff of $1/\Delta\tau$ from $10$ to $20$ respectively) results in a small change in the non-universal numerical value of $\lambda_s$ (which defines the boundary between the strange metal phase and the glassy SRO phase, and is analogous to a QCP in a conventional quantum critical theory) from $5.0$ to $5.125$ respectively. However, importantly, physical results at $\lambda = \lambda_s$ remain unchanged (Fig. \ref{fig:trotter_comparison}).

\begin{figure*}
    \centering
    \includegraphics[width=0.98\textwidth]{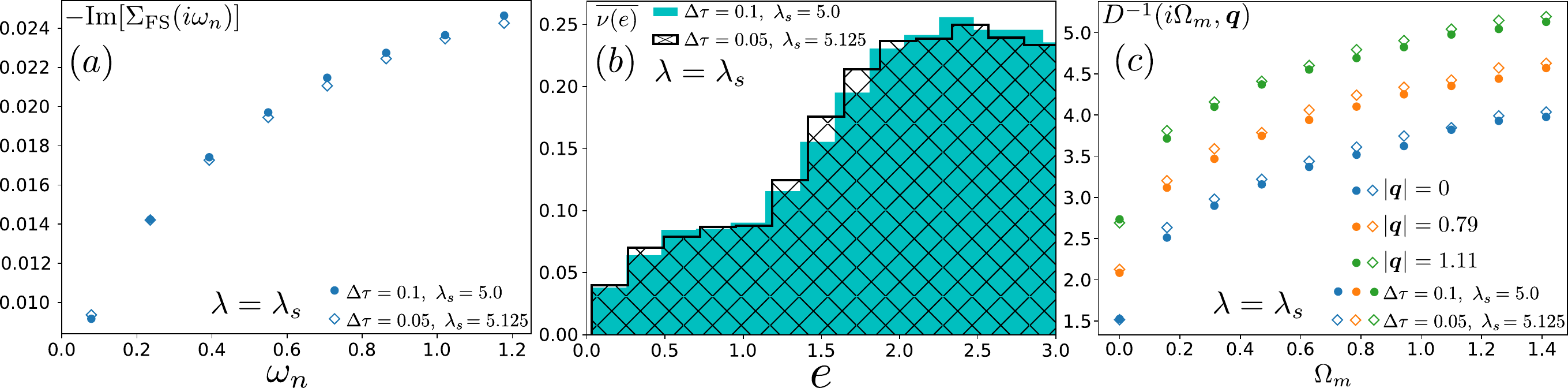}
    \caption{Comparison of physical results at $\lambda = \lambda_s$ and inverse temperature $\beta = 40$ for two different Trotter step sizes, $\Delta\tau = 0.1$ and $\Delta\tau = 0.05$: (a) The imaginary part of the Matsubara fermion self-energy averaged over the FS. (b) The disorder-averaged boson density of states. (c) The inverse disorder-averaged boson propagator in momentum space. The results are nearly identical. There is a small difference in the nonuniversal numerical value of the parameter $\lambda_s$ defining the boundary between the strange metal phase and the glassy SRO phase.}
    \label{fig:trotter_comparison}
\end{figure*}

\end{itemize}

\end{document}